\documentclass[journal]{IEEEtran}
\usepackage[T1]{fontenc}
\usepackage[latin9]{inputenc}
\usepackage{color}
\usepackage{float}
\usepackage{bm}
\usepackage{amsmath}
\usepackage{amsthm}
\usepackage{amssymb}
\usepackage{stackrel}
\usepackage{graphicx}
\usepackage[unicode=true,
 bookmarks=false,
 breaklinks=false,pdfborder={0 0 0},pdfborderstyle={},backref=false,colorlinks=false]
 {hyperref}
\hypersetup{pdftitle={Your Title},
 pdfauthor={Your Name},
 pdfpagelayout=OneColumn, pdfnewwindow=true, pdfstartview=XYZ, plainpages=false}

\makeatletter

\floatstyle{ruled}
\newfloat{algorithm}{tbp}{loa}
\providecommand{\algorithmname}{Algorithm}
\floatname{algorithm}{\protect\algorithmname}

\theoremstyle{plain}
\newtheorem{lem}{\protect\lemmaname}
\theoremstyle{plain}
\newtheorem{thm}{\protect\theoremname}
\theoremstyle{plain}
\newtheorem{cor}{\protect\corollaryname}

\usepackage[caption=false,font=footnotesize]{subfig}
\usepackage{algorithmic}
\usepackage{algorithm}

\usepackage{bm}
\usepackage{cite}
\usepackage[english]{babel}
\usepackage[T1]{fontenc}

\IEEEoverridecommandlockouts

\@ifundefined{showcaptionsetup}{}{%
 \PassOptionsToPackage{caption=false}{subfig}}
\usepackage{subfig}
\makeatother

\providecommand{\corollaryname}{Corollary}
\providecommand{\lemmaname}{Lemma}
\providecommand{\theoremname}{Theorem}

\begin{document}
\title{Age of Changed Information: Content-Aware Status Updating in the Internet
of Things}
\author{Xijun~Wang,~\IEEEmembership{Member,~IEEE,} Wenrui~Lin, Chao Xu,~\IEEEmembership{Member,~IEEE,}
Xinghua~Sun,~\IEEEmembership{Member,~IEEE,}  and Xiang~Chen,~\IEEEmembership{Member,~IEEE}\thanks{Part of this work was presented at the IEEE WCNC, May 2020 \cite{linAverageAgeChanged2020a}.
This work was supported in part by the State\textquoteright s Key
Project of Research and Development Plan under Grant 2019YFE0196400,
in part by Guangdong R\&D Project in Key Areas under Grant 2019B010158001,
in part by Guangdong Basic and Applied Basic Research Foundation under
Grants 2021A1515012631 and 2019A1515011906, in part by Key Laboratory
of Modern Measurement \& Control Technology, Ministry of Education,
Beijing Information Science \& Technology University (KF20201123202),
in part by the National Natural Science Foundation of China (61701372),
in part by the Chinese Universities Scientific Fund (2452017560),
High-level Talents Fund of Shaanxi Province (F2020221001), Technological
Innovation Fund of Shaanxi Academy of Forestry (SXLK2021-0215). (\emph{Corresponding
authors: Chao Xu and Xiang Chen})}\thanks{X. Wang and X. Chen is with School of Electronics and Information
Technology, Sun Yat-sen University, Guangzhou, China (e-mail: wangxijun@mail.sysu.edu.cn;
chenxiang@mail.sysu.edu.cn). }\thanks{W. Lin and X. Sun are with School of Electronics and Communication
Engineering, Sun Yat-sen University, Guangzhou, China (e-mail: linwr7@mail2.sysu.edu.cn;
sunxinghua@mail.sysu.edu.cn).}\thanks{C. Xu is with School of Information Engineering, Northwest A\&F University,
Yangling, Shaanxi, China (e-mail: cxu@nwafu.edu.cn). }\thanks{Copyright (c) 2021 IEEE. Personal use of this material is permitted.
However, permission to use this material for any other purposes must
be obtained from the IEEE by sending a request to pubs-permissions@ieee.org.}}
\maketitle
\begin{abstract}
In Internet of Things (IoT), the freshness of status updates is crucial
for mission-critical applications. In this regard, it is suggested
to quantify the freshness of updates by using Age of Information (AoI)
from the receiver's perspective. Specifically, the AoI measures the
freshness over time. However, the freshness in the content is neglected.
In this paper, we introduce an age-based utility, named as \emph{Age
of Changed Information} (AoCI), which captures both the passage of
time and the change of information content. By modeling the underlying
physical process as a discrete time Markov chain, we investigate the
AoCI in a time-slotted status update system, where a sensor samples
the physical process and transmits the update packets to the destination.
With the aim of minimizing the weighted sum of the AoCI and the update
cost, we formulate an infinite horizon average cost Markov Decision
Process. We show that the optimal updating policy has a special structure
 with respect to the AoCI and identify the condition under which
the special structure exists. By exploiting the special structure,
we provide a low complexity relative policy iteration algorithm that
finds the optimal updating policy. We further investigate the optimal
policy for two special cases. In the first case where the state of
the physical process transits with equiprobability, we show that optimal
policy is of threshold type and derive the closed-form of the optimal
threshold. We then study a more generalized periodic Markov model
of the physical process in the second case. Lastly, simulation results
are laid out to exhibit the performance of the optimal updating policy
and its superiority over the zero-wait baseline policy. 
\end{abstract}

\begin{IEEEkeywords}
Internet of things, information freshness, Markov decision processes,
structural analysis.
\end{IEEEkeywords}

\section{Introduction}

With the sharp proliferation of the Internet of Thing (IoT) devices
and the rising need of mission-critical services, timely delivery
of information has become increasingly important in real-time status
update systems, such as environmental monitoring in smart city, vehicle
tracking in autonomous driving, and video surveillance in smart home,
whose performance strongly depends on the freshness of the status
updates received by the destination \cite{palattellaInternetThings5G2016,schulzLatencyCriticalIoT2017,abd-elmagidRoleAgeofInformationInternet2019}.
The Age of Information (AoI) has been recently introduced to measure
information freshness from the receiver\textquoteright s perspective
\cite{kaulRealtimeStatusHow2012}. Particularly, it is defined as
the time elapsed since the generation of the most recent status update
packet received by the destination. In general, the smaller the AoI
at the destination, the fresher the received status update. The AoI
jointly characterizes the packet delay and the packet intergeneration
time, which distinguishes AoI from conventional metrics, such as delay
and throughput. Such a superiority of AoI for evaluating the information
freshness in various wireless networks has been demonstrated in recent
studies \cite{kaulRealtimeStatusHow2012,kaulStatusUpdatesQueues2012,kostaAgeInformationPerformance2018,xuOptimizingInformationFreshness2020}
by resorting to queueing theory.

An IoT network mainly consists of three components, i.e. the IoT device,
the communication network, and the destination node. As such, to optimize
the information freshness in terms of the AoI for the IoT, it is of
great importance to control the status update process, which has attracted
significant research attention in recent years \cite{hsuSchedulingAlgorithmsMinimizing2020,sunUpdateWaitHow2017,bedewyAgeoptimalSamplingTransmission2018,liuUAVAidedDataCollection2020,ceranAverageAgeInformation2019,tangMinimizingAgeInformation2020,zhouJointStatusSampling2019,lengAgeInformationMinimization2019a,abd-elmagidReinforcementLearningFramework2020,abd-elmagidAoIOptimalJointSampling2020}.
Particularly, authors in \cite{hsuSchedulingAlgorithmsMinimizing2020}
studied the AoI optimal packet transmission policy with random information
arrivals under a transmission capacity constraint. In the case where
the IoT device generates status updates at will, the authors in \cite{sunUpdateWaitHow2017}
studied the AoI minimization problem with a single device, where a
zero-wait policy was shown to be non-optimal. Minimizing the average
AoI in a status update system with multiple devices was further studied
in \cite{bedewyAgeoptimalSamplingTransmission2018}. Two age-optimal
data collection problems were investigated to minimize the average
AoI and peak AoI for UAV enabled wireless sensor networks in \cite{liuUAVAidedDataCollection2020}.
An optimal status updating scheme with hybrid Automatic Repeat request
(ARQ) was proposed in \cite{ceranAverageAgeInformation2019} to minimize
the average AoI under a constraint on the average number of transmissions.
By considering the restriction from bandwidth and power consumption
constraints, a dynamic scheduling algorithms was developed to minimize
the average AoI of industrial IoT networks in \cite{tangMinimizingAgeInformation2020}.
The authors in \cite{zhouJointStatusSampling2019} designed a joint
status sampling and updating to minimize the average AoI under an
average energy cost constraint. Further, by empowering the sensor
nodes with energy harvesting techniques, recent work \cite{lengAgeInformationMinimization2019a}
developed the age-aware primary spectrum sensing and update strategy
for an energy harvesting cognitive radio. Meanwhile, the wireless
energy transfer procedure and scheduling of update packet transmissions
were jointly optimized by authors in \cite{abd-elmagidReinforcementLearningFramework2020},
who further extended their research by considering the sampling cost
in \cite{abd-elmagidAoIOptimalJointSampling2020}.

As seen above, the AoI has been widely used as a performance metric
to characterize the information freshness over time. It does, however,
disregard the content carried by the updates and the current knowledge
at the receiver. A natural question that then emerges is whether measuring
the freshness of updates through the AoI alone is sufficient. Several
recent attempts have been made to answer this question \cite{sunSamplingDataFreshness2019,kamEffectiveAgeInformation2018a,zhongTwoFreshnessMetrics2018,tangSchedulingMinimizeAge2020,maatoukAgeIncorrectInformation2020}.
The pros and cons of these metrics are elaborated in the following.
\begin{itemize}
\item Authors in \cite{sunSamplingDataFreshness2019} utilized mutual information
between the state of the physical process and the received updates
at the destination to evaluate the information freshness. The mutual
information quantifies the amount of information that the received
updates carry about the current value of the physical process. Although
the destination has no knowledge of the current value of the physical
process, it is proved that the mutual information is a non-negative
and non-increasing function of AoI if the physical process is a stationary
Markov chain and the sampling times are independent of the value of
the physical process. Therefore, the mutual information can be computed
by the destination via the AoI. However, for more general cases where
the sampling policy needs to be devised based on the causal knowledge
of the value of the physical process, the mutual information is not
necessarily a function of the age. In this light, how to compute the
mutual information at the destination is unknown.
\item In \cite{kamEffectiveAgeInformation2018a}, the authors proposed a
metric, called sampling age, which is the time difference between
the last ideal sampling time and the first actual sampling time. The
ideal sampling time is the most recent time at which the state of
the physical process changed relative to the last received update.
However, the ideal sampling time is not available to the destination
and hence the sampling age cannot be obtained by the destination.
\item The Age of Synchronization (AoS) was proposed in \cite{zhongTwoFreshnessMetrics2018}
to measure how long the information at the receiver has become desynchronized
compared with the physical process. It is defined as the time difference
between the current time and the earliest update generation time after
the previous synchronization time \cite{tangSchedulingMinimizeAge2020}.
Similar to the AoI, the AoS drops when the destination receives a
status update packet. However, unlike the AoI which begins to increase
immediately after the reception of a status update packet at the destination,
the AoS remains to be zero and does not increase until the sensor
generates a new status update packet. However, the destination does
not know the generation time of the earliest update after the previous
synchronization until it receives this update. Hence, the AoS cannot
be calculated at the destination.
\item The Age of Incorrect Information (AoII) was proposed in \cite{maatoukAgeIncorrectInformation2020}
to address the real-time remote estimation problem. The AoII combines
a time penalty function and an estimation error penalty function that
reflects the difference between the current estimate at the destination
and the actual state of the physical process. As such, the AoII will
increase with time when the receiver stays in an erroneous state.
Computing the AoII at the destination requires that the state of the
physical process is available to the destination in any time slot.
Otherwise, the estimation at the destination and the state of the
physical process cannot be compared to compute the estimation error
penalty function in the AoII. However, the destination cannot observe
the state of the physical process until it receives the status update
packet. Therefore, the AoII cannot be computed by the destination.
\end{itemize}

In summary, mutual information cannot be computed if the sampling
times are determined by using causal knowledge of the value of the
physical process, while the other three metrics are only available
to the transmitter rather than the destination. Since the last three
metrics cannot be computed by the destination, they cannot be applied
in the scenario where the sensor has no computing capability and the
destination is in charge of decision-making. This is also the scenario
we focus on in this work. Moreover, even if the sensor has the computing
capability, the above three metrics require the continuous sensing
of the physical process, which would induce noticeable energy consumption.

In this paper, we concentrate on the scenario where the receiver aims
to conduct timely detection of status changes in the underlying physical
process only based on its received update packets. In practice, a
status change won\textquoteright t be detected until an update generated
after the change point is successfully delivered to the destination
for the first time. However, it is impossible to know the exact time
instant of a status change unless the physical process is monitored
continuously. Therefore, it is challenging to design the optimal updating
policy to balance the information freshness and the energy consumption.
On the one hand, sampling and transmitting at a higher frequency incurs
a higher energy consumption of the sensor. On the other hand, sampling
at a lower frequency results in staleness in detecting a status change
or even a miss detection. The error-prone wireless channel further
worsens the situation, since the update packet may be dropped due
to channel outage. As a result, the receiver could be fooled into
believing that no change in state has taken place. Motivated by all
this, we introduce a utility function from the receiver's perspective
that depicts both the passage of time and the change of information
content. We further investigate this utility in a status update system
consisting of a sensor and a destination. In particular, the sensor
monitors the real-time status of a physical process, which is modeled
by a discrete time Markov chain with uniform stationary distribution,
and transmits status update packets to the destination. In our earlier
work \cite{linAverageAgeChanged2020a}, we investigated the effects
of content change on the information freshness and designed the optimal
status updating policy in an IoT system. However, the model of the
physical process is limited to the two-state Markov chain with the
equal transition probabilities. The key contributions of this paper
are summarized as follows:
\begin{itemize}
\item Motivated by the fact that a status change will not be perceived by
the destination until an update generated after the change instant
is successfully delivered, we introduce a new age-based utility,
referred to as Age of Changed Information (AoCI), that characterizes
the information freshness via the updates received by the destination.
The word ``changed'' refers to the newly received update that brings
new content different from the previous one at the destination. The
AoCI takes into account the information content of the updates and
the current knowledge at the destination. It will increase when the
update with the same status information is received. 
\item We formulate the status updating problem as an infinite horizon average
cost Markov Decision Process (MDP) with the goal of minimizing the
weighted sum of the AoCI and the update cost. By incorporating the
AoCI into the cost function of the MDP, the sensor is made to sample
and transmit at a higher frequency when the same status information
is continuously received, thereby potentially reducing the miss detection.
We analyze the properties of the value function without specifying
the state transition model of the physical process. Armed with these
properties, we show that the optimal updating policy has a special
structure with respect to the AoCI and identify the condition on the
return probability of the physical process under which the special
structure exists. A structure-aware relative policy iteration algorithm
is then proposed to obtain the optimal updating policy with low complexity. 
\item We study two special cases, where the return probability satisfies
the condition.  In the first case, by giving an example that the
state of the underlying physical process transits with equiprobability,
we simplify the MDP and prove that the optimal policy is of threshold
type. We also derive the optimal threshold in closed-form, which sheds
insight on how the system parameters affect the threshold policy.
Particularly, we prove that the optimal threshold is non-increasing
with transmission success probability and the number of states of
the physical process, but is non-decreasing with the update cost.
We generalize the example in the first case by studying the periodic
Markov model of the physical process in the second case. Simulation
results highlight interesting insights on the effects of the system
parameters and show the superiority of the optimal updating policy
over the zero-wait policy.
\end{itemize}

The rest of the paper is organized as follows: Section II provides
a description of the system model and a definition for the proposed
performance metric. In Section III, we present the MDP formulation
and analyze the structure of the optimal policy. Two special cases
are then analyzed in Section IV. Simulation results are presented
in Section V, followed by the conclusion in Section VI.

\section{System Overview \label{sec:System-Overview}}

\subsection{System Model}

\begin{figure}[tp]
\centering

\includegraphics[width=0.5\textwidth]{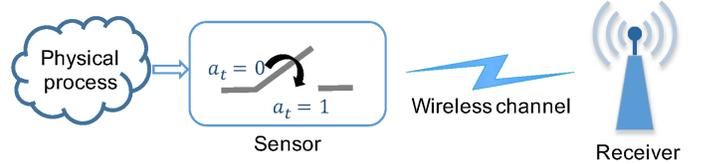}\caption{\label{fig:SystemModel}An illustration of a status update system
monitoring a physical process.}
\end{figure}
As shown in Fig. \ref{fig:SystemModel}, we consider a status update
system consisting of a sensor and a destination, where the sensor
performs simple monitoring tasks, such as reading temperature, and
the destination could be a monitor or an actuator. The status update
system is assumed to be time-slotted. The sensor may produce a status
update about the underlying time-varying process (a.k.a. generate-at-will)
in each time slot and send it over an unreliable wireless channel
to the destination. The sensor could also remain idle in one time
slot. Let $a_{t}\in\{0,1\}$ be the action of the sensor in the $t$-th
slot, where $a_{t}=1$ indicates that the sensor samples and transmits
an update, and $a_{t}=0$, otherwise. Any update's transmission time
is assumed to be equal to one slot length. Moreover, the slot length
is normalized to unity, without loss of generality. In general, each
update would entail a cost $C_{u}$, including the sampling cost and
the transmission cost, where the cost for sampling is assume to be
negligible compared to that of transmission. 

Assume that the underlying time-varying physical process is modeled
by a $M$-state discrete time Markov chain $\{X_{t};t\in\mathbb{N}\}$
with $X_{t}\in\{1,2,...,M\}$, where $M\geq2$. The period of each
state is equal to the length of one slot, and the transition occurs
at the beginning of each slot just before the sampling decision. We
assume that all the state of the Markov chain have have uniform stationary
distribution.

We assume that the fading of channels in each slot stays constant
but varies independently over various slots. We also assume that an
update is transmitted by the sensor at a fixed rate, and channel state
information is only available at destination. As such, the transmission
in each time slot may fail because of an outage and thus the loss
of the packet could be described by a memoryless Bernoulli process.
Specifically, let $h_{t}\in\{0,1\}$ denote whether the transmission
succeeds or fails, where $h_{t}=1$ indicates that the transmission
is successful, and $h_{t}=0$, otherwise. We define the success probability
as $\Pr\{h_{t}=1\}=p_{s}$ and the failure probability as $\Pr\{h_{t}=0\}=p_{f}=1-p_{s}$.
After the destination receives the update packet, a single-bit acknowledgement
is fed back instantaneously without error. We assume that the failed
update will be discarded, and if the sensor decides to transmit in
the next slot, a new status update will be generated.\footnote{The reason why we sample and transmit a new status update rather than
retransmit a failed update lies in two aspects. On the one hand,
according to the studies in \cite{ceranAverageAgeInformation2019},
in the context of AoI, it is better not to retransmit an undecoded
packet with the classical ARQ protocol, where failed transmissions
are discarded at the destination and the receiver tries to decode
each retransmission as a new message. This is because the probability
of a successful transmission is the same for a retransmission and
for the transmission of a new update. On the other hand, in this work
we focus on the scenario that the sensor performs simple monitoring
tasks, such as reading temperature, and hence, the cost for generating
status packets is assumed to be negligible compared to that of transmission.
Then, the energy costs for transmitting a new update and retransmitting
a failed update are almost the same. Altogether, we choose to transmit
a new update when the transmission failure occurs.}  

\subsection{Freshness Metric}

We assume that at the beginning of a slot a status update is generated
and transmitted, and the destination will receive it at the end of
the slot if the transmission succeeds. The AoI, commonly used to quantify
the freshness of the information, is specified as the time elapsed
since the generation of the latest status update received by the destination.
Suppose that the latest status update successfully received by the
destination was generated at the time instants $U(t)$, i.e., $U(t)=\max\{g_{i}\mid d_{i}\leq t\}$,
where $g_{i}$ and $d_{i}$ represent the time instants when the update
$i$ is generated and delivered, respectively. Then, the AoI at the
beginning of slot t is given by
\begin{equation}
\delta_{t}=t-U(t).
\end{equation}

The proposed metric, AoCI, is different from the AoI in that the AoCI
not only captures the time lag of the update received at the destination,
but also includes variations in the information content of these updates.
In particular, the AoCI only declines when the newly received update
content differs from the previous one, and boosts otherwise. Let us
denote by $n(t)=\max\{i|d_{i}\leq t\}$ the index of the latest update
the destination receives at the end of slot $t$. We let $Y_{j}$
denote the information content of update $j$. It is worth noting
that $Y_{j}$ is equal to the state of the physical process in the
slot when update $j$ was generated, e.g., $Y_{n(t)}=X_{U(t)}$. Let
us denote by $m(t)=\max\{j|Y_{j}\neq Y_{n(t)},d_{j}\leq d_{n(t)}\}$
the index of the most recently update which has different content
from the latest update $n(t)$ got. Then, we can define the AoCI at
the beginning of slot $t$ as
\begin{equation}
\Delta_{t}=t-U'(t),\label{eq:AoCI}
\end{equation}
where $U'(t)=\min\{g_{k}|d_{m(t)}<d_{k}\leq d_{n(t)}\}$ represents
the generation time of the next successfully received update after
$m(t)$. Noting that all the update packets that have been successfully
received after $m(t)$ have the same content as the latest one received.
We set the upper limits to the AoCI and the AoI, which are denoted
by $\hat{\Delta}$ and $\hat{\delta}$, respectively.

\begin{figure}[t]
\centering

\includegraphics[width=0.5\textwidth]{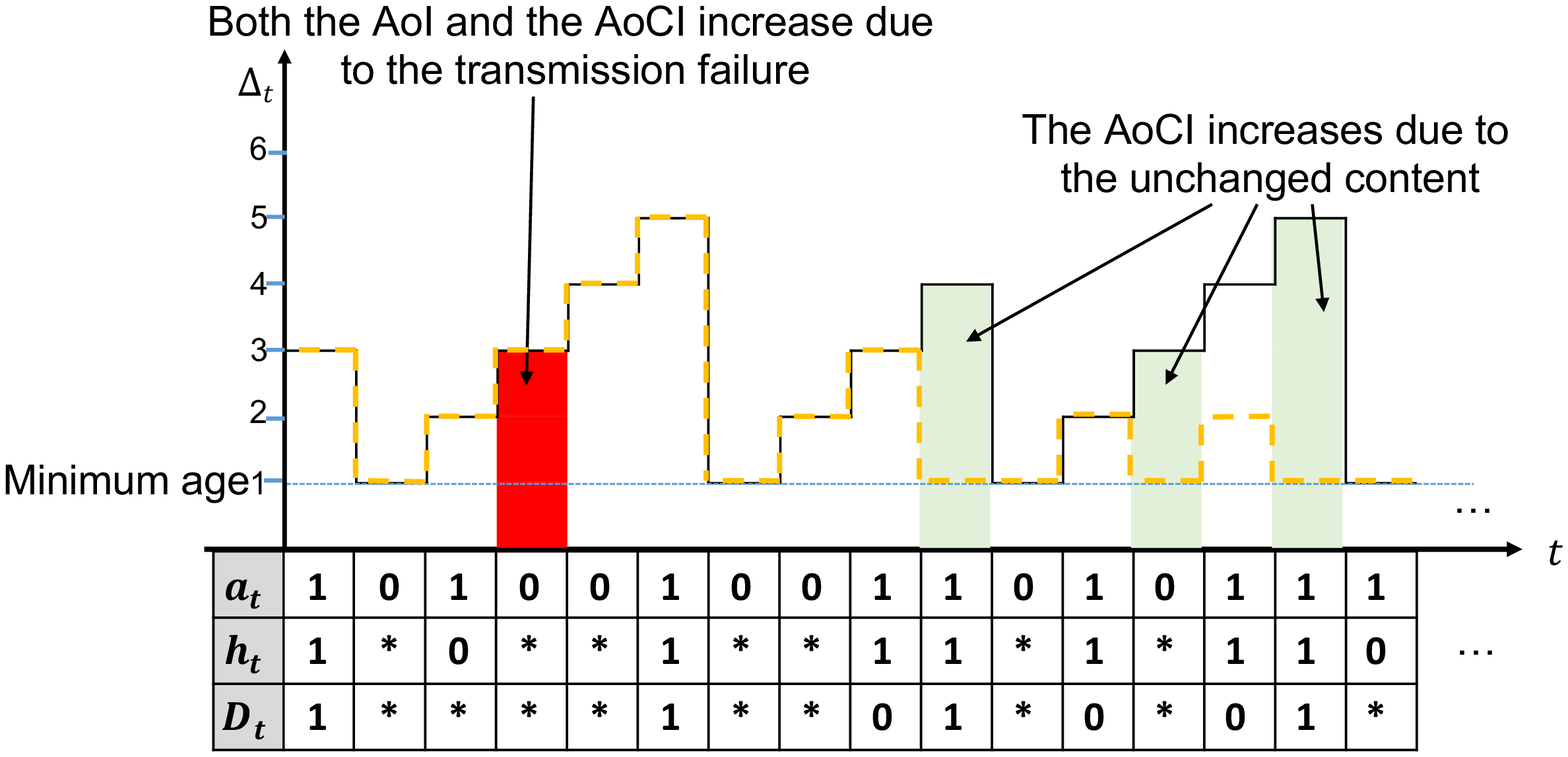}\caption{\label{fig:AOI figure}An illustration of the AoCI and the AoI in
a time-slotted status update system.}
\end{figure}

In slot $t$ where a status update is received successfully, we denote
by $D_{t}\in\{0,1\}$ an indicator for whether the content of the
newly received update varies from that of the previously received
update. If $D_{t}=1$, then the newly received update has different
content. Otherwise, it has the same content. Particularly, the content
change probability is defined as 
\begin{align}
\Pr(D_{t}=1) & \overset{(a)}{=}\Pr(Y_{n(t)}\neq Y_{n(t)-1})\overset{(b)}{=}1-p_{r}(\delta_{t}),
\end{align}
where $p_{r}(\delta_{t})=\Pr(X_{U(t)}=X_{U(t)-\delta_{t}})$ is the
return probability that the state of the physical process remains
the same after $\delta_{t}$ steps. In the above equation, (a) holds
due to the definition of $D_{t}$, and (b) holds because of the fact
that $Y_{n(t)}=X_{U(t)}$ and $Y_{n(t)-1}=X_{U(t)-\delta_{t}}$.\footnote{Note that the time difference between two consecutive samples at the
sensor is $\delta_{t}$. Therefore, the update $n(t)-1$ was generated
at $U(t)-\delta_{t}$, and its content $Y_{n(t)-1}$ is the same as
the state of the physical process at $U(t)-\delta_{t}$, i.e., $X_{U(t)-\delta_{t}}$. } According to (\ref{eq:AoCI}), if a new status update generated by
the sensor is received successfully by the destination (i.e., $a_{t}=1,h_{t}=1$)
and it contains different content from the update previously received
(i.e., $D_{t}=1$), then the AoCI decreases to one; otherwise, the
AoCI increases by one. Then, the dynamics of the AoCI is given by
\begin{equation}
\Delta_{t+1}=\begin{cases}
1, & a_{t}=1,h_{t}=1,D_{t}=1;\\
\min\{\Delta_{t}+1,\hat{\Delta}\}, & \text{otherwise}.
\end{cases}\label{eq:Dynamic}
\end{equation}
For ease of exposition, we demonstrate how the AoCI and the AoI evolve
over time in Fig. \ref{fig:AOI figure}, where the dotted line represents
the AoI and the solid line represents the AoCI.

\subsection{Problem Formulation}

The aim of this paper is to find an update policy $\pi=(a_{0},a_{1},\ldots)$
that minimizes the total average cost, which is defined as the weighted
sum of the AoCI and the update cost. By defining $\Pi$ as a set of
stationary and deterministic policies,\footnote{A policy is said to be stationary and deterministic if it is time
invariant and chooses an action with probability one.} our problem is formulated as follows:
\begin{equation}
\min_{\pi\in\Pi}\limsup_{T\rightarrow\infty}\frac{1}{T}\stackrel[t=0]{T}{\sum}\mathbb{E}[\Delta_{t}+\omega a_{t}C_{u}],\label{eq:Problem}
\end{equation}
where $\omega$ is the weighting factor and the expectation is taken
with respect to the distribution over trajectories induced by $\pi$
together with the transition probabilities.

\textcolor{red}{}
\begin{figure*}[tb]
\begin{equation}
\begin{cases}
\Pr(\bm{s}_{t+1}=(\min\{\Delta+1,\hat{\Delta}\},\min\{\delta+1,\hat{\delta}\})|\bm{s}_{t}=(\Delta,\delta),a_{t}=0)=1,\\
\Pr(\bm{s}_{t+1}=(\min\{\Delta+1,\hat{\Delta}\},\min\{\delta+1,\hat{\delta}\})|\bm{s}_{t}=(\Delta,\delta),a_{t}=1)=p_{f},\\
\Pr(\bm{s}_{t+1}=(\min\{\Delta+1,\hat{\Delta}\},1)|\bm{s}_{t}=(\Delta,\delta),a_{t}=1)=p_{s}p_{r}(\delta),\\
\Pr(\bm{s}_{t+1}=(1,1)|\bm{s}_{t}=(\Delta,\delta),a_{t}=1)=p_{s}(1-p_{r}(\delta)),
\end{cases}\label{eq:state-transition}
\end{equation}
\hrulefill
\end{figure*}

\section{Optimal Update Policy Design \label{sec:Updating-Policy-Design}}

\subsection{MDP Formulation}

The optimization problem in (\ref{eq:Problem}) can be cast into an
infinite horizon average cost Markov decision process $(\mathcal{S},\mathcal{A},\Pr(\cdot|\cdot,\cdot),C(\cdot,\cdot))$,
where each element is described as follows:
\begin{itemize}
\item States: The state of the MDP at time slot $t$ is defined to be the
tuple of the AoCI and the AoI, i.e., $\bm{s}_{t}\triangleq\left(\Delta_{t},\delta_{t}\right)$.
Since both the AoCI and the AoI are bounded by their upper limits,
the state space $\mathcal{S}$ is finite.
\item Actions: The action at time slot $t$ is $a_{t}$ and the action set
is $\mathcal{A}=\{0,1\}$. 
\item Transition Probability: Let $\Pr(\bm{s}_{t+1}|\bm{s}_{t},a_{t})$
denote the transition probability that state transits from $\bm{s}_{t}$
to $\bm{s}_{t+1}$ by taking action $a_{t}$ at slot $t$. Because
the event of packet transmission and that of content change are independent,
according to the AoCI evolution dynamics (\ref{eq:Dynamic}), the
transition probability is represented as in (\ref{eq:state-transition})
and $\Pr(\bm{s}_{t+1}|\bm{s}_{t},a_{t})=0$ otherwise. 
\item Cost: We let $C(\bm{s}_{t},a_{t})=\Delta_{t}+\omega a_{t}C_{u}$ denote
the instantaneous cost at state $\bm{s}_{t}$ given action $a_{t}$.
\end{itemize}

The above MDP is a finite-state finite-action average-cost MDP. According
to \cite[Theorem 8.4.5]{putermanMarkovDecisionProcesses2005}, there
exists a deterministic stationary average optimal policy for the finite-state
finite-action average-cost MDP if the cost function is bounded and
the MDP is unichain, i.e., the Markov chain corresponding to every
deterministic stationary policy consists of a single recurrent class
plus a possibly empty set of transient states. Below, we examine these
two conditions. First, the cost of the above MDP is bounded since
the instantaneous cost is defined as the weighted sum of the AoCI
and the energy consumption. Second, since the state $(\hat{\Delta},\hat{\delta})$
is reachable from all other states, the induced Markov chain has a
single recurrent class. Hence, the MDP is unchain. Altogether, there
exists a stationary and deterministic optimal policy. The optimal
policy $\pi^{*}$ to minimize the total average cost can be obtained
by solving the following Bellman equation \cite{dimitrip.bertsekasDynamicProgrammingOptimal2007}:
\begin{equation}
\theta+V(\bm{s})=\min_{a\in\{0,1\}}\left\{ C(\bm{s},a)+\sum_{\bm{s}'\in\mathcal{S}}\Pr(\bm{s}'|\bm{s},a)V(\bm{s}')\right\} ,\forall\bm{s}\in\mathcal{S},\label{eq:Bellman}
\end{equation}
where $\theta$ is the optimal value to (\ref{eq:Problem}) for all
initial state and $V(\bm{s})$ is the value function which is a mapping
from $\bm{s}$ to real values. Moreover, for any $\bm{s}\in\mathcal{S}$,
the optimal policy can be given by
\begin{equation}
\pi^{*}(\bm{s})=\arg\min_{a\in\{0,1\}}\left\{ C(\bm{s},a)+\sum_{\bm{s}'\in\mathcal{S}}\Pr(\bm{s}'|\bm{s},a)V(\bm{s}')\right\} .\label{eq:optimal-policy}
\end{equation}
We can seen from (\ref{eq:optimal-policy}) that the optimal policy
$\pi^{*}$ depends on the value function $V(\cdot)$. Unfortunately,
there is usually no closed-form solution for $V(\cdot)$ \cite{dimitrip.bertsekasDynamicProgrammingOptimal2007}.
Therefore, numerous numerical algorithms have been proposed in the
literature, such as policy iteration and value iteration. Nonetheless,
owing to the curse of dimensionality these approaches are typically
computationally exhausting, and few insights can be leveraged for
optimal policy. Hence, in the sequel, we study the structural properties
of the optimal updating policy. 

To analyze the structure of $\pi^{*}$, we introduce the state-action
value function $Q(\bm{s},a)$, which is defined as 
\begin{equation}
Q(\bm{s},a)=C(\bm{s},a)+\sum_{\bm{s}'\in\mathcal{S}}\Pr(\bm{s}'|\bm{s},a)V(\bm{s}'),
\end{equation}
for all $\bm{s}\in\mathcal{S}$ and $a\in\mathcal{A}$. Note that
$Q(\bm{s},a)$ is related to the RHS of the Bellman equation in (\ref{eq:Bellman}).
The optimal policy can also be expressed in terms of $Q(\bm{s},a)$,
i.e., 
\begin{equation}
\pi^{*}(\bm{s})=\arg\min_{a\in\{0,1\}}Q(\bm{s},a),\quad\forall\bm{s}\in\mathcal{S}.\label{eq:optimal-policy-Q}
\end{equation}

\subsection{Structural Analysis and Optimal Policy}

Before we present the main theorem, we first show the key properties
of the value function $V(\Delta,\delta)$ in the following lemmas. 
\begin{lem}
\label{lem:value function monotony wrt AOCI}The value function $V(\Delta,\delta)$
is non-decreasing with $\Delta$ for any $\delta$.
\end{lem}
\begin{IEEEproof}
See Appendix \ref{subsec:Proof v-function AoCI}.
\end{IEEEproof}
\begin{lem}
\label{lem:slope wrt aoci}Given $\delta$, we have $V(\Delta_{2},\delta)-V(\Delta_{1},\delta)\geq\Delta_{2}-\Delta_{1}$
for any $\Delta_{2}\geq\Delta_{1}$. 
\end{lem}
\begin{IEEEproof}
See Appendix \ref{subsec:Proof-of-slope wrt aoci}.
\end{IEEEproof}

\begin{lem}
\label{lem:slope wrt aoi}For any $\Delta$ and $\delta_{1}\leq\delta_{2}$,
if $p_{r}(\delta_{1})-p_{r}(\delta_{2})\leq\frac{\delta_{2}-\delta_{1}}{\mathop{V_{k}(\Delta+1,1)}-V_{k}(1,1)}$
for any $k\in\mathbb{Z}_{\geq0}$, we have $V(\Delta,\delta_{1})-V(\Delta,\delta_{2})\leq\delta_{2}-\delta_{1}$,
where $V_{k}(\cdot)$ is the value function obtained in the value
iteration algorithm.
\end{lem}
\begin{IEEEproof}
See Appendix \ref{subsec:Proof-of-slope wrt aoi}.
\end{IEEEproof}
Then, we present the structure of the optimal updating policy in the
following theorem.
\begin{thm}
\label{thm:threshold-structure-general}Given $\bm{s}=(\Delta,\delta)$,
if $p_{r}(1)-p_{r}(\delta+1)\leq\frac{\delta}{\mathop{V_{k}(\Delta+1,1)}-V_{k}(1,1)}$
for any $\Delta$ and $k\in\mathbb{Z}_{\geq0}$, the optimal updating
policy has a special structure for any $\delta$, that is, if $\Delta\geq\underline{\Delta}(\delta)$,
then $\pi^{*}(\bm{s})=1$, where $\underline{\Delta}(\delta)$ is
the minimum integer value satisfying $p_{s}(1-p_{r}(\delta))\Delta-p_{s}\delta-\omega C_{u}\geq0$.
\end{thm}
\begin{IEEEproof}
See Appendix \ref{subsec:Proof-of-threshold-structure}.
\end{IEEEproof}
It is noteworthy that the special structure in Theorem 1 is different
from the threshold structure defined in available studies \cite{zhouJointStatusSampling2019,maatoukAgeIncorrectInformation2020},
where the optimal policy is to update when $\Delta$ is no less than
the threshold and the optimal policy is to keep idle otherwise.

We then propose a low-complexity relative policy iteration algorithm
to compute the optimal policy based on the special structure of the
optimal updating policy presented in Theorem \ref{thm:threshold-structure-general}.
Although the exact value of the threshold is not available in the
general case, we can still reduce the computational complexity for
obtaining the optimal policy by exploiting the special structure.
This is because the threshold structure only relies on the properties
of the value function. Particularly, if $p_{r}(1)-p_{r}(\delta+1)\leq\frac{\delta}{\mathop{V_{k}(\hat{\Delta},1)}-V_{k}(1,1)}$
and $p_{s}(1-p_{r}(\delta))\Delta-p_{s}\delta-\omega C_{u}\geq0$
hold,\footnote{According to Lemma 1, $V_{k}(\hat{\Delta},1)\geq V_{k}(\Delta,1)$
for $1\leq\Delta<\hat{\Delta}$. Therefore, if $p_{r}(1)-p_{r}(\delta+1)\leq\frac{\delta}{\mathop{V_{k}(\hat{\Delta},1)}-V_{k}(1,1)}$,
then $p_{r}(1)-p_{r}(\delta+1)\leq\frac{\delta}{\mathop{V_{k}(\Delta+1,1)}-V_{k}(1,1)}$
for any $\Delta$.} then it is optimal to update for the state $(\Delta,\delta)$. Therefore,
we can determine the optimal action immediately (Lines 6-7 in Algorithm
1). Otherwise, we have to perform the minimization to find the optimal
action (Lines 8-9 in Algorithm 1). Since Algorithm 1 is a monotone
policy iteration algorithm, the policy $\pi_{k}(\bm{s})$ and value
function $V_{k}(\bm{s})$ will finally converge when $k$ increases.
The details of the proposed relative policy iteration algorithm is
summarized in Algorithm 1. By taking the advantage of the special
structure of the optimal policy, the complexity of the policy improvement
step in the relative policy iteration algorithm can be reduced. Since
the upper limits of the AoCI and the AoI are $\hat{\Delta}$ and $\hat{\delta}$,
respectively, the cardinality of the state space is $\left|\hat{\Delta}\right|\times\left|\hat{\delta}\right|$.
According to \cite{littmanComplexitySolvingMarkov1995}, the computational
complexity saving for each iteration in the structure-aware policy
iteration algorithm is $O\left(\left(\left|\hat{\Delta}\right|\times\left|\hat{\delta}\right|\right)^{2}\right)$. 

\begin{algorithm}[tb]
\caption{Relative Policy Iteration based on the Threshold Structure}

\begin{algorithmic}[1]

\STATE Initialization: Set $k=0$ and $\pi_{0}(\bm{s})=0$ for all
state $\bm{s}=(\Delta,\delta)\in S$, select a reference state $\bm{s}^{\dagger}$
and set $V_{0}(\bm{s}^{\dagger})=0$.

\REPEAT

\STATE $\pi_{k+1}(\bm{s})\leftarrow0$.

\medskip{}

Policy Evaluation:

\STATE Given policy $\pi_{k}(\bm{s})$, compute the value of $\theta_{k}$
and $V_{k}(\bm{s})$ by solving the following $|S|$ linear equations:

$\begin{cases}
\theta_{k}+V_{k}(\bm{s})=C(\bm{s},\pi_{k}(\bm{s}))+\\
\qquad\sum\limits _{\bm{s}'\in\mathcal{S}}\Pr(\bm{s}'|\bm{s},\pi_{k}(\bm{s}))V_{k}(\bm{s}'),\\
V_{k}(\bm{s}^{\dagger})=0.
\end{cases}$

\medskip{}

Policy Improvement:

\FOR{$\bm{s}=(\Delta,\delta)\in S$}

\IF{$p_{r}(1)-p_{r}(\delta+1)\leq\frac{\delta}{\mathop{V_{k}(\hat{\Delta},1)}-V_{k}(1,1)}$
and $p_{s}(1-p_{r}(\delta))\Delta-p_{s}\delta-\omega C_{u}\geq0$}

\STATE $\pi_{k+1}(\bm{s})\leftarrow1$.

\ELSE

\STATE $\pi_{k+1}(\bm{s})\leftarrow\arg\min\limits _{a\in\mathcal{A}}\{C(\bm{s},a)+\sum\limits _{\bm{s}'\in\mathcal{S}}\Pr(\bm{s}'|\bm{s},a)V_{a}(\bm{s}')\}$.

\ENDIF

\ENDFOR

\STATE $k\leftarrow k+1$.

\UNTIL{$\pi_{k+1}(\bm{s})=\pi_{k}(\bm{s})$ for all $\bm{s}\in\mathcal{S}$.}

\STATE $\pi^{*}\leftarrow\pi_{k+1}$.

\RETURN the optimal policy $\pi^{*}$.

\end{algorithmic}
\end{algorithm}

\section{Special Case Study\label{sec:Special-Case-Study}}

In this section, we study two special cases, where the return probability
of the physical process satisfies certain conditions.

\subsection{Case 1}

We first consider a special case where the return probability $p_{r}(\delta)$
is irrespective of $\delta$, i.e., $p_{r}(\delta_{1})=p_{r}(\delta_{2})\neq0$
for any $\delta_{1}\neq\delta_{2}$. It is easy to see that the condition
in Theorem \ref{thm:threshold-structure-general} is satisfied in
this special case and the optimal updating policy has a threshold
structure with respect to the AoCI. 

One example for this special case is that the state of the underlying
physical process transits with equiprobability. The one-step state
transition probability matrix for $M$-state discrete time Markov
chain is given by
\begin{equation}
\Pr(X_{t+1}|X_{t})=\left[\begin{array}{cccc}
p_{c} & p_{c} & \ldots & p_{c}\\
p_{c} & p_{c} & \ldots & p_{c}\\
\vdots & \vdots & \ddots & \vdots\\
p_{c} & p_{c} & \ldots & p_{c}
\end{array}\right],\label{eq:transition-matrix-2}
\end{equation}
where $p_{c}=\frac{1}{M}$. Since the $\delta$-step state transition
probability matrix $\Pr(X_{t+\delta}|X_{t})$ is the same with $\Pr(X_{t+1}|X_{t})$
for all $\delta$, we have $p_{r}(\delta)=1/M$. 

In this case, we can simplify the MDP formulated in Section \ref{sec:Updating-Policy-Design}.A.
In particular, the state at slot $t$ is only the AoCI, i.e., $s_{t}=\Delta_{t}$,
and the state transition probability in (\ref{eq:state-transition})
can be simplified as 
\begin{equation}
\begin{cases}
\Pr(s_{t+1}=\min\{\Delta+1,\hat{\Delta}\}|s_{t}=\Delta,a_{t}=0)=1,\\
\Pr(s_{t+1}=\min\{\Delta+1,\hat{\Delta}\}|s_{t}=\Delta,a_{t}=1)=p_{f}+p_{s}p_{r},\\
\Pr(s_{t+1}=1|s_{t}=\Delta,a_{t}=1)=p_{s}(1-p_{r}),
\end{cases}
\end{equation}
and $\Pr(s_{t+1}|s_{t},a_{t})=0$ otherwise. 

\begin{figure}[tp]
\centering

\includegraphics[width=0.5\textwidth]{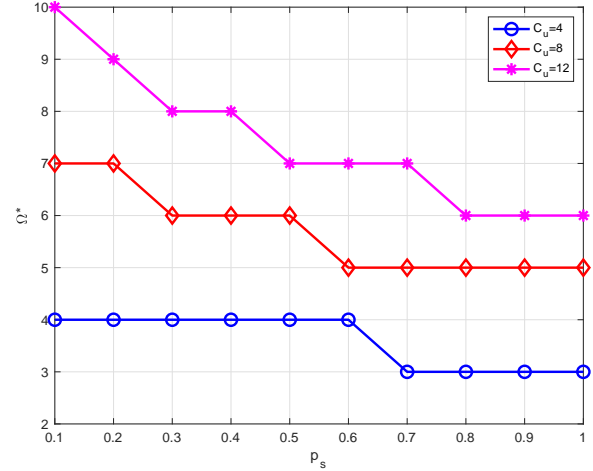}\caption{\label{fig:Optimal threshold}The asymptotically optimal threshold
$\Omega^{*}$ versus $p_{s}$ ($M=2$, $p_{c}=0.5$, and $\omega=1$).}
\end{figure}

Based on the simplified state and transition probability, we present
the monotonicity property of $V(s)$ in the following lemma. 
\begin{lem}
\label{lem:lemma2}The value function $V(s)$ is non-decreasing with
$s$. 
\end{lem}
\begin{IEEEproof}
See Appendix \ref{subsec:Proof-of-Lemma 2}.
\end{IEEEproof}
Next, in the following theorem, we give results on the structure of
the optimal updating policy.
\begin{thm}
\label{thm:threshold type}The optimal policy has a threshold structure,
that is, if $\pi^{*}(s_{1})=1$, then $\pi^{*}(s_{2})=1$ for all
$s_{2}\geq s_{1}$.
\end{thm}
\begin{IEEEproof}
See Appendix \ref{subsec:Proof-of-Theorem-1}.\emph{}
\end{IEEEproof}
According to Theorem \ref{thm:threshold type}, the optimal policy
can be represented as a threshold policy, which is given by
\begin{equation}
\pi^{*}(s)=\begin{cases}
1, & \text{if }s\ge\Omega,\\
0, & \text{otherwise},
\end{cases}\label{eq:Threshold}
\end{equation}
where $\Omega$ is the threshold at which the switching occurs. Under
the threshold policy, we proceed with analyzing the total average
cost of any threshold $\Omega$ in the asymptotic regime.
\begin{lem}
\label{lem:total average cost}Let $p_{z}\triangleq p_{f}+p_{s}p_{r}$.
When $\hat{\Delta}$ goes to infinity, for any given threshold $\Omega$,
the total average cost $J(\Omega)$ of the threshold policy approaches
to $J(\Omega)=J_{1}(\Omega)+J_{2}(\Omega)$, where 
\begin{align}
J_{1}(\Omega)= & \frac{1-p_{z}}{\Omega(1-p_{z})+p_{z}}\left(\frac{\Omega(\Omega-1)}{2}+\frac{\Omega}{1-p_{z}}+\frac{p_{z}}{(1-p_{z})^{2}}\right),
\end{align}
and
\begin{equation}
J_{2}(\Omega)=\frac{\omega C_{u}}{\Omega(1-p_{z})+p_{z}}.
\end{equation}
\end{lem}
\begin{IEEEproof}
See Appendix \ref{subsec:Proof-of-expected value of average cost}.
\end{IEEEproof}
By leveraging the above results, we can proceed to find the optimal
threshold value $\Omega^{*}$.
\begin{thm}
\label{thm:closed-form} The asymptotically optimal threshold $\Omega^{*}$
of the optimal updating policy is given by
\begin{equation}
\Omega^{*}=\arg\min(J(\left\lfloor \Omega'\right\rfloor ),J(\left\lceil \Omega'\right\rceil )),\label{eq:optimalThreshold}
\end{equation}
where $\Omega'=\frac{\sqrt{p_{z}+2\omega C_{u}(1-p_{z})}-p_{z}}{1-p_{z}}.$
\end{thm}
\begin{IEEEproof}
See Appendix \ref{subsec:Proof-of-closed-form}.
\end{IEEEproof}
\begin{cor}
\label{cor:optimal-threshold}The asymptotically optimal threshold
$\Omega^{*}$ is non-decreasing with the update cost $C_{u}$, but
is non-increasing with transmission success probability $p_{s}$ and
the number of states of the physical process $M$.
\end{cor}
\begin{IEEEproof}
See Appendix \ref{subsec:Proof-of-Corollary1}.
\end{IEEEproof}
Fig. \ref{fig:Optimal threshold} illustrates the asymptotically optimal
threshold $\Omega^{*}$ of the optimal updating policy with respect
to $p_{s}$ under different $C_{u}$. The asymptotically optimal threshold
is shown to be non-increasing with $p_{s}$. This is because, before
the destination successfully receives an update packet, the sensor
has to sample and transmit more times when $p_{s}$ is small. Hence,
updating the status is productive only when the AoCI is large. We
can also observe that the asymptotically optimal threshold is non-decreasing
with $C_{u}$. This indicates that the sensor will remain idle until
the AoCI is large, if the update cost is high. Hence, the optimal
policy is able to achieve a balance between the AoCI and the update
cost.

\begin{figure}[tp]
\centering

\includegraphics[width=0.5\textwidth]{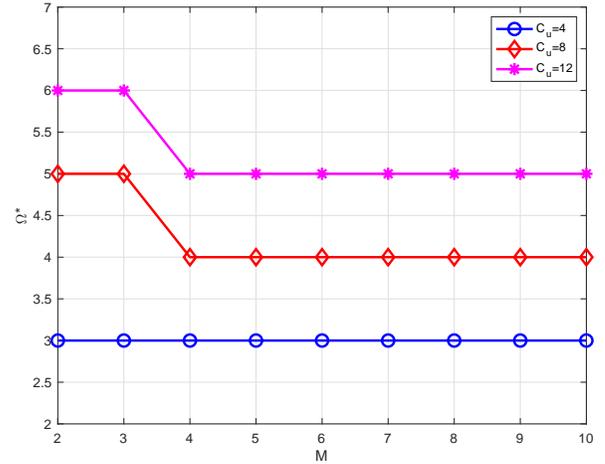}\caption{\label{fig:Optimal threshold withMps=00003D1}The asymptotically optimal
threshold $\Omega^{*}$ versus $M$ ($p_{c}=1/M$, $p_{s}=1$, and
$\omega=1$).}
\end{figure}

Fig. \ref{fig:Optimal threshold withMps=00003D1} shows the asymptotically
optimal threshold $\Omega^{*}$ of the optimal updating policy with
respect to $M$ under different $C_{u}$. We can observe that the
asymptotically optimal threshold is non-increasing with $M$. This
is due to the fact that the return probability $p_{r}$ is large,
when $M$ is small. In other words, the received status update is
more likely to contain the same content with the previous one. Hence,
it is more cost-efficient to have a larger threshold at a smaller
$M$. We note that the reason why $\Omega^{*}$ becomes constant when
$M$ becomes large is due to the fact that $\Omega'$ converges as
$M$ grows and so do $\left\lfloor \Omega'\right\rfloor $ and $\left\lceil \Omega'\right\rceil $. 

\subsection{Case 2 }

In this case, we consider that $p_{r}(1)=0$ and $p_{r}(\delta)\geq0$
for $\delta>1$. It is also easy to see that the condition in Theorem
\ref{thm:threshold-structure-general} is satisfied in this special
case and the optimal updating policy has a special structure with
respect to the AoCI as given in Theorem \ref{thm:threshold-structure-general}.
The optimal updating policy can also be obtained via Algorithm 1.

The example for this case is the physical process modeled by periodic
Markov chain. For instance, the Markov model could be a $M$-state
one-dimensional random walk with the state transition probability
matrix given by 
\begin{equation}
\Pr(X_{t+1}|X_{t})=\left[\begin{array}{cccccc}
0 & p_{c} & 0 & \ldots & 0 & 1-p_{c}\\
1-p_{c} & 0 & p_{c} & \ldots & 0 & 0\\
\vdots & \vdots & \vdots & \ddots &  & \vdots\\
p_{c} & 0 & 0 & \ldots & 1-p_{c} & 0
\end{array}\right],\label{eq:transition-matrix-3}
\end{equation}
where $p_{c}\in(0,1)$ and $M$ is even. Every state of this Markov
chain is periodic with period 2. Hence, the return probability $p_{r}(\delta)$
is $0$ when $\delta$ is odd, and is positive when $\delta$ is even.

\begin{figure}[tp]
\centering

\includegraphics[width=0.5\textwidth]{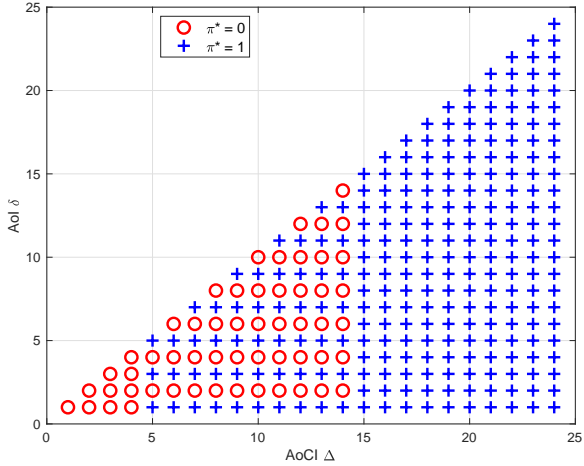}\caption{\label{fig:Structure of optimal policy-1}Structure of the optimal
policy for different values of $\delta$ for one-dimensional random
walk ($M=4$, $p_{c}=0.5$, $p_{s}=0.8$, $C_{u}=12$, and $\omega=1$).}
\end{figure}

In Figs. \ref{fig:Structure of optimal policy-1} and \ref{fig:Structure of optimal policy pc(1/M,1/M-1)},
we illustrate the analytical results of Theorem \ref{thm:threshold-structure-general}
when Markov model of the physical process is a one-dimensional random
walk with 4-state and 6-state, respectively. The period of both Markov
chain is 2. In both figures, the optimal updating policy is shown
to have a special structure with respect to the AoCI for any $\delta$.
In fact, the structure of the optimal policy unveils a tradeoff between
the AoCI and the update cost. Particularly, if the AoCI is small,
it is not efficient for the sensor to send the status update to the
destination due to the update cost. It can also be seen that the optimal
action $\pi^{*}=0$ does not appear in the whole state space of the
AoCI if the AoI is high. This is due to the fact that the AoCI is
always no less than the AoI and it is more efficient to generate a
new status update due to the outdated information at the destination
with the high AoI. 

\begin{figure}[tp]
\centering

\includegraphics[width=0.5\textwidth]{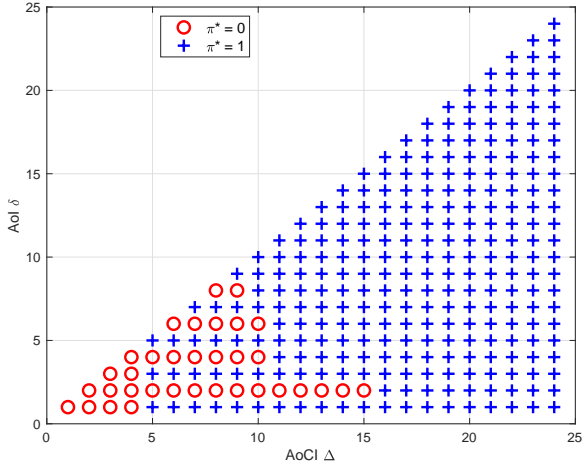}\caption{\label{fig:Structure of optimal policy pc(1/M,1/M-1)}Structure of
the optimal policy for different values of $\delta$ for one-dimensional
random walk ($M=6$, $p_{c}=0.5$, $p_{s}=0.8$, $C_{u}=12$, and
$\omega=1$).}
\end{figure}

\section{Simulation Results \label{sec:Simulation-Results}}

In this section, the simulation results of the optimal updating policy
are presented to examine the effects of system parameters. The performance
of the optimal updating policy is also compared with that of two baseline
policies, i.e., zero-wait policy and sample-at-change policy. In zero-wait
policy, the sensor samples and transmits the status update at each
time slot. While in sample-at-change policy, a new update is generated
only when the state changes relative to the previous received update
at the destination. Note that the sample-at-change policy is genie-aided
and can achieve the minimum AoCI.

\subsection{Performance Evaluation in the Special Case 1}

\begin{figure}[tb]
\centering\subfloat[]{\centering

\includegraphics[width=0.5\textwidth]{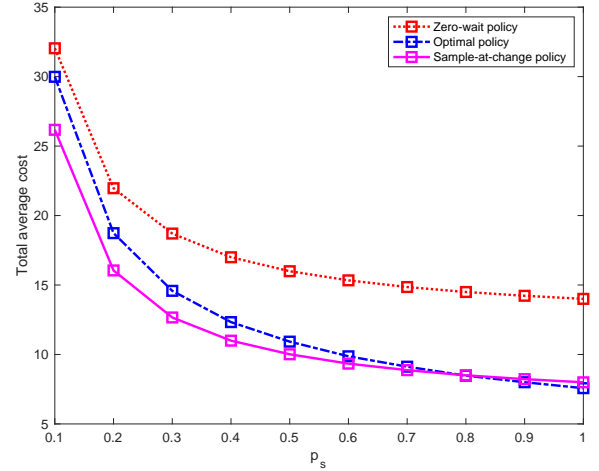}}

\subfloat[]{\centering

\includegraphics[width=0.5\textwidth]{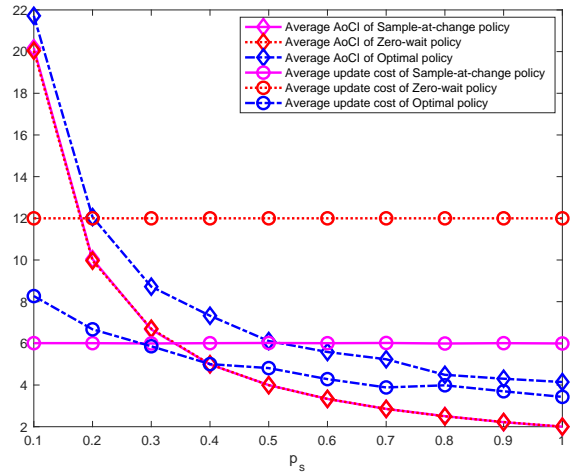}}

\caption{\label{fig:Comparision-ps}Comparison between the optimal policy,
sample-at-change policy, and zero-wait policy ($M=2$, $p_{c}=0.5$,
$C_{u}=12$ and $\omega=1$). (a) The total average cost versus $p_{s}$.
(b) The average AoCI and the average update cost versus $p_{s}$.}
\end{figure}

In Fig. \ref{fig:Comparision-ps}, we compare the total average cost
of the optimal policy and two baseline policies with respect to $p_{s}$.
It can be observed in Fig. \ref{fig:Comparision-ps}\,(a) that the
optimal policy outperforms the zero-wait policy. Moreover, as $p_{s}$
increases, there is a larger reduction in the total average cost.
The reason can be explained with the aid of Fig. \ref{fig:Comparision-ps}\,(b).
Although the zero-wait policy gains a smaller AoCI, it bears a constant
update cost. In contrast, the optimal policy can trade off the AoCI
for the update cost. Particular, in the optimal policy, the sensor
remains idle until the AoCI is larger than a threshold, thereby inducing
a large AoCI. However, the optimal policy has a smaller update cost
than the zero-wait policy. We also observe that the sample-at-change
policy outperforms the optimal policy at first. As $p_{s}$ increases,
the optimal policy beats the sample-at-change policy, because the
update cost of the optimal policy declines as $p_{s}$ increases.
Hence, the optimal policy is more cost-efficient. 

\begin{figure}[tb]
\centering\subfloat[]{\centering

\includegraphics[width=0.5\textwidth]{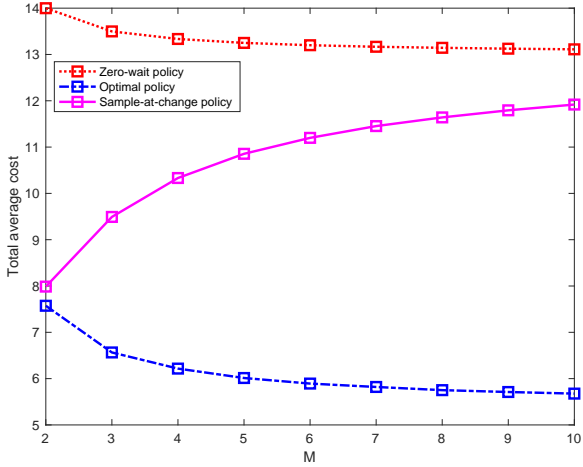}}

\subfloat[]{\centering

\includegraphics[width=0.5\textwidth]{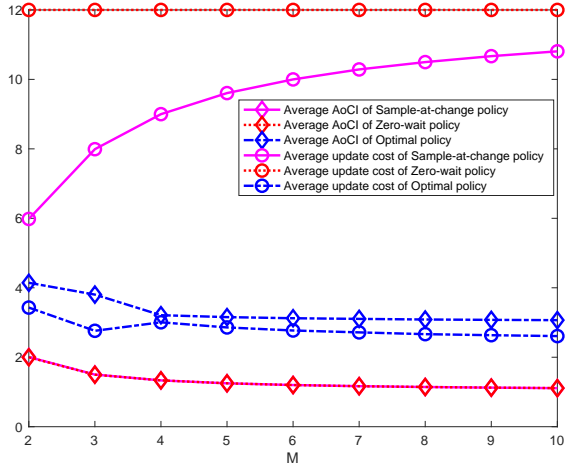}}

\caption{\label{fig:Comparision bewteen op and zwp different M pc=00003D1/M-1}Comparison
between the optimal policy, sample-at-change policy, and zero-wait
policy ($p_{c}=1/M$, $p_{s}=1$, $C_{u}=12$ and $\omega=1$). (a)
The total average cost versus $M$. (b) The average AoCI and average
update cost versus $M$.}
\end{figure}

In  Fig. \ref{fig:Comparision bewteen op and zwp different M pc=00003D1/M-1},
we compare the total average cost of the optimal policy and two baseline
policies with respect to $M$, where $p_{c}$ is set to be $1/M$.
We can see in Fig. \ref{fig:Comparision bewteen op and zwp different M pc=00003D1/M-1}\,(a)
that the optimal policy outperforms both baseline policies. Moreover,
for both the optimal policy and the zero-wait policy, when $M$ grows,
the total average cost decreases. As shown in Fig. \ref{fig:Comparision bewteen op and zwp different M pc=00003D1/M-1}\,(b),
the average AoCI of the optimal policy is larger than that of the
zero-wait policy, whereas the average update cost of the optimal policy
is smaller than that of the zero-wait policy. The average AoCI decreases
with the increasing of $M$ because the received status update is
more likely to have a new content with a large $M$. The average update
cost is affected by $M$ and the optimal threshold simultaneously.
The increase of the average update cost when $M=4$ is due to the
decrease of the optimal threshold as shown in Fig. \ref{fig:Optimal threshold withMps=00003D1}.
When the optimal threshold is fixed, the average update cost decreases
as $M$ increases. We also observe that the total average cost of
the sample-at-change policy increases with $M$. This is because the
return probability decreases with $M$ and the sample-at-change policy
approaches the zero-wait policy as $M$ increases. 

\subsection{Performance Evaluation in the Special Case 2}

\begin{figure}[tb]
\centering\subfloat[]{\centering

\includegraphics[width=0.5\textwidth]{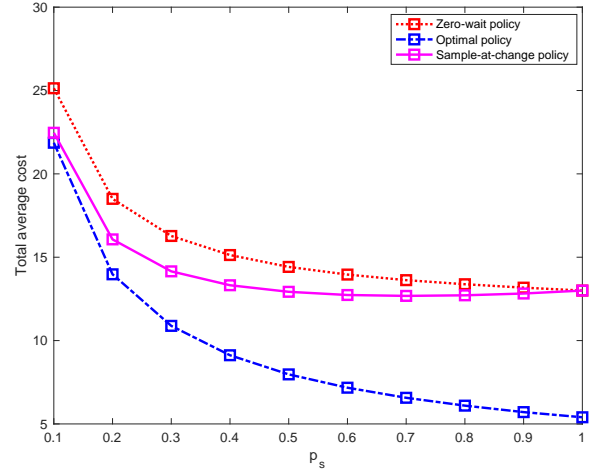}}

\subfloat[]{\centering

\includegraphics[width=0.5\textwidth]{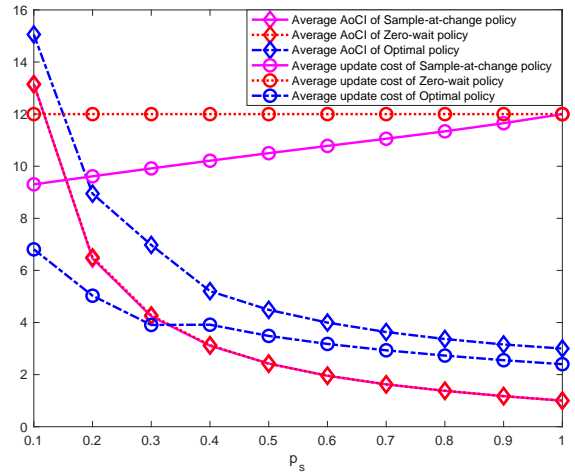}}

\caption{\label{fig:Comparision-ps-2}Comparison between the optimal policy,
sample-at-change policy, and zero-wait policy ($M=4$, $p_{c}=0.2$,
$C_{u}=12$ and $\omega=1$). (a) The total average cost versus $p_{s}$.
(b) The average AoCI and the average update cost versus $p_{s}$.}
\end{figure}

In Fig. \ref{fig:Comparision-ps-2}, we compare the total average
cost of the optimal policy and two baseline policies with respect
to $p_{s}$. It can be observed that the optimal policy outperforms
both two baseline policies. As $p_{s}$ increases, both the AoCI and
the update cost of the optimal policy decline. This is because the
AoCI can be reset with fewer transmission when $p_{s}$ is large.
The AoCI of both baseline policy decreases with the increase of $p_{s}$.
However, the update cost of the zero-wait policy remains constant
and the update cost of the sample-at-change policy increases with
$p_{s}$. In particular, the sample-at-change policy degenerates to
zero-wait policy when $p_{s}=1$. As a result, the performance gain
of the optimal policy is larger when $p_{s}$ is larger. 

\begin{figure}[tb]
\centering\subfloat[]{\centering

\includegraphics[width=0.5\textwidth]{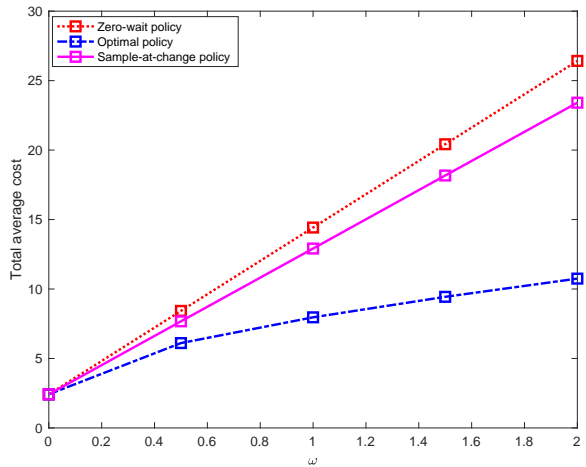}}

\subfloat[]{\centering

\includegraphics[width=0.5\textwidth]{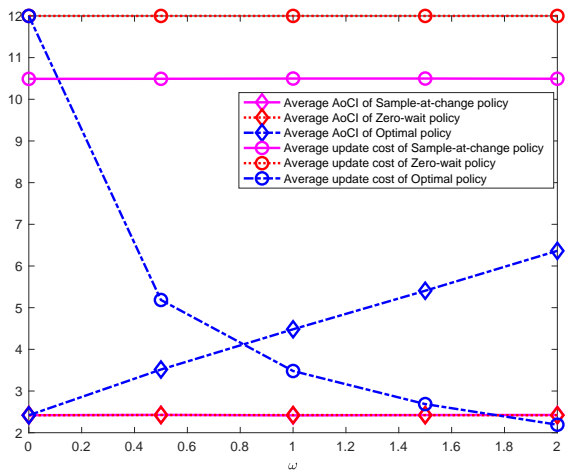}}

\caption{\label{fig:Comparision bewteen op and zwp different w}Comparison
between the optimal policy, sample-at-change policy, and zero-wait
policy ($M=4$, $p_{c}=0.2$, $p_{s}=0.5$ and $C_{u}=12$ ). (a)
The total average cost versus $\omega$. (b) The average AoCI and
average update cost versus $\omega$.}
\end{figure}

In Fig. \ref{fig:Comparision bewteen op and zwp different w}, we
compare the total average cost of the optimal policy and two baseline
policies with respect to $\omega$. From Fig. \ref{fig:Comparision bewteen op and zwp different w}\,(a),
we can observe that the total average cost of three policies increase
with the weighting factor $\omega$. The three policies are coincident
when $\omega=0$, which implies that the sensor under the optimal
policy would sample and transmit in each time slot. As $\omega$ increases,
the total average cost of the zero-wait policy and the sample-at-change
policy grow linearly. This is because these two baseline policies
cannot adapt to the weighting factor and its average AoCI and average
update cost are constant, as shown in Fig. \ref{fig:Comparision bewteen op and zwp different w}\,(b).
On the contrary, the optimal policy is able to adjust according to
$\omega$. When $\omega$ grows larger, the AoCI is traded off to
obtain a smaller update cost. Therefore, the optimal policy can strike
a balance between the AoCI and the update cost.

\section{Conclusion\label{sec:Conclusion}}

In this paper, by identifying the ignorance of information content
variation in the conventional AoI, we have proposed the AoCI as an
age-based utility to quantify information freshness. The AoCI not
only measures the freshness by the passage of time but also captures
the information content of the updates at the destination. We have
further investigated the updating policy in the status update system
by taking into account both the AoCI and the update cost, and formulated
the status updating problem as an infinite horizon average cost MDP.
We have analyze the properties of the value function without specifying
the state transition model of the physical process. Based on these
properties, we have proved that the optimal updating policy has a
special structure with respect to the AoCI and identify the condition
on the return probability of the physical process under which the
special structure exists. Equipped with this, we have provided a structure-aware
relative policy iteration algorithm to obtain the optimal updating
policy with low complexity. We have also studied two special cases
where the condition holds. In the special case where the state of
the underlying physical process transits with equiprobability, we
have proved that the optimal policy is of threshold type and derived
the closed-form of the optimal threshold. We have also proved that
the optimal threshold is non-increasing with respect to transmission
success probability and the number of states of the physical process,
respectively, but is non-decreasing with respect to the update cost.
Results from the simulation have shown the impacts of the unreliable
channel and the physical process on the total average cost. By comparing
the optimal updating policy with the zero-wait policy and the sample-at-change
policy, it is shown that the optimal updating policy achieves a balance
between the AoCI and the update cost and yields a substantial performance
boost in terms of the total average cost compared to zero-wait policy.

\appendix{}

\subsection{Proof of Lemma \ref{lem:value function monotony wrt AOCI}\label{subsec:Proof v-function AoCI}}

We prove Lemma \ref{lem:value function monotony wrt AOCI} through
mathematical induction and the value iteration algorithm (VIA) \cite{dimitrip.bertsekasDynamicProgrammingOptimal2007}.
We first briefly introduce VIA. For each state $\bm{s}$, let $V_{k}(\bm{s})$
be the value function at iteration $k$. In VIA, the value function
can be updated as follows:
\begin{equation}
V_{k+1}(\bm{s})=\min_{a}\left\{ Q_{k}(\bm{s},a)\right\} ,\forall\bm{s}\in\mathcal{S}.
\end{equation}
Under any initialization of the initial value $V_{0}(\bm{s})$, the
sequence $\{V_{k}(\bm{s})\}$ converges to the value function in the
Bellman equation (\ref{eq:Bellman}) \cite{dimitrip.bertsekasDynamicProgrammingOptimal2007},
i.e.,
\begin{equation}
\lim_{k\rightarrow\infty}V_{k}(\bm{s})=V(\bm{s}),\forall s\in\mathcal{S}.\label{eq:lim-value-function}
\end{equation}
Therefore, the monotonicity of $V(\bm{s})$ in $\mathcal{S}$ can
be guaranteed by proving that for any $\bm{s}_{1},\bm{s}_{2}\in\mathcal{S}$,
such that $\bm{s}_{1}\leq\bm{s}_{2}$,
\begin{equation}
V_{k}(\bm{s}_{1})\leq V_{k}(\bm{s}_{2}),\quad k=0,1,\ldots\label{eq:monotonicity}
\end{equation}

According to (\ref{eq:lim-value-function}), the monotonicity of $V(\bm{s})$
with respect to the AoCI can be guaranteed by proving that for any
$\bm{s}_{1}=(\Delta_{1},\delta),\bm{s}_{2}=(\Delta_{2},\delta)\in\mathcal{S}$,
such that $\Delta_{1}\leq\Delta_{2}$,
\begin{equation}
V_{k}(\bm{s}_{1})\leq V_{k}(\bm{s}_{2}),\quad k=0,1,\ldots\label{eq:monotonicity-2}
\end{equation}

Then, we prove (\ref{eq:monotonicity-2}) via mathematical induction.
Without loss of generality, we initialize $V_{0}(\bm{s})=0$ for all
$\bm{s}\in\mathcal{S}$. Thus, (\ref{eq:monotonicity-2}) holds for
$k=0$. Next, we assume that (\ref{eq:monotonicity-2}) holds up till
$k>0$ and we examine whether it holds for $k+1$. 

When $a=0$, we have $Q_{k}(\bm{s}_{1},0)=\Delta_{1}+\mathop{V_{k}(\bm{s}'_{1})}$
and $Q_{k}(\bm{s}_{2},0)=\Delta_{2}+\mathop{V_{k}(\bm{s}'_{2})}$,
where $\bm{s}'_{1}=(\Delta_{1}+1,\delta+1)$ and $\bm{s}'_{2}=(\Delta_{2}+1,\delta+1)$.
Since $\Delta_{1}+1\leq\Delta_{2}+1$ and $V_{k}(\bm{s}'_{1})\leq V_{k}(\bm{s}'_{2})$,
we can easily see that $Q_{k}(\bm{s}_{1},0)\leq Q_{k}(\bm{s}_{2},0)$.

When $a=1$, we have 
\begin{align}
Q_{k}(\bm{s}_{1},1)= & \Delta_{1}+\omega C_{u}+p_{s}(1-p_{r}(\delta))V_{k}(1,1)\nonumber \\
 & +p_{f}\mathop{V_{k}(\Delta_{1}+1,\delta+1)}+p_{s}p_{r}(\delta)\mathop{V_{k}(\Delta_{1}+1,1)}
\end{align}
 and 
\begin{align}
Q_{k}(\bm{s}_{2},1)= & \Delta_{2}+\omega C_{u}+p_{s}(1-p_{r}(\delta))V_{k}(1,1)\nonumber \\
 & +p_{f}\mathop{V_{k}(\Delta_{2}+1,\delta+1)}+p_{s}p_{r}(\delta)\mathop{V_{k}(\Delta_{2}+1,1)}.
\end{align}
 Since $V_{k}(\Delta_{1},\delta)\leq V_{k}(\Delta_{2},\delta)$ for
any $\delta$, we can also verify that $Q_{k}(\bm{s}_{1},1)\leq Q_{k}(\bm{s}_{2},1)$.

Altogether, we can assert that $V_{k+1}(s_{1})\leq V_{k+1}(s_{2})$
for any $k$. By taking limits on both sides of (\ref{eq:monotonicity-2})
and by (\ref{eq:lim-value-function}), we complete the proof of Lemma
\ref{lem:value function monotony wrt AOCI}. 

\subsection{Proof of Lemma \ref{lem:slope wrt aoci}\label{subsec:Proof-of-slope wrt aoci}}

Let $\bm{s}_{1}=(\Delta_{1},\delta)$ and $\bm{s}_{2}=(\Delta_{2},\delta)$.
Based on Lemma \ref{lem:value function monotony wrt AOCI}, we have
\begin{align}
Q(\bm{s}_{2},0)-(\Delta_{2}-\Delta_{1})= & \Delta_{1}+V(\Delta_{2}+1,\delta+1)\nonumber \\
\geq & \Delta_{1}+V(\Delta_{1}+1,\delta+1)=Q(\bm{s}_{1},0),
\end{align}
and
\begin{align}
 & Q(\bm{s}_{2},1)-(\Delta_{2}-\Delta_{1})\nonumber \\
= & \Delta_{1}+\omega C_{u}+p_{s}(1-p_{r}(\delta))V(1,1)+p_{f}\mathop{V(\Delta_{2}+1,\delta+1)}\nonumber \\
 & +p_{s}p_{r}(\delta)\mathop{V(\Delta_{2}+1,1)}\nonumber \\
\geq & \Delta_{1}+\omega C_{u}+p_{s}(1-p_{r}(\delta))V(1,1)+p_{f}\mathop{V(\Delta_{1}+1,\delta+1)}\nonumber \\
 & +p_{s}p_{r}(\delta)\mathop{V(\Delta_{1}+1,1)}\nonumber \\
= & Q(\bm{s}_{1},1).
\end{align}

Since $V({\bf s})=\underset{a}{\min}Q({\bf s},a)$, we can prove $V(\bm{s}_{2})-V(\bm{s}_{1})\geq\Delta_{2}-\Delta_{1}$
in four cases as follows:
\begin{itemize}
\item Case 1: If $V(\bm{s}_{1})=Q(\bm{s}_{1},0)$ and $V(\bm{s}_{2})=Q(\bm{s}_{2},0)$,
then $V(\bm{s}_{2})-V(\bm{s}_{1})=Q(\bm{s}_{2},0)-Q(\bm{s}_{1},0)\geq\Delta_{2}-\Delta_{1}$.
\item Case 2: If $V(\bm{s}_{1})=Q(\bm{s}_{1},1)$ and $V(\bm{s}_{2})=Q(\bm{s}_{2},1)$,
then $V(\bm{s}_{2})-V(\bm{s}_{1})=Q(\bm{s}_{2},1)-Q(\bm{s}_{1},1)\geq\Delta_{2}-\Delta_{1}$.
\item Case 3: If $V(\bm{s}_{1})=Q(\bm{s}_{1},0)\leq Q(\bm{s}_{1},1)$ and
$V(\bm{s}_{2})=Q(\bm{s}_{2},1)$, then $V(\bm{s}_{2})-V(\bm{s}_{1})=Q(\bm{s}_{2},1)-Q(\bm{s}_{1},0)\geq Q(\bm{s}_{2},1)-Q(\bm{s}_{1},1)\geq\Delta_{2}-\Delta_{1}$.
\item Case 4: If $V(\bm{s}_{1})=Q(\bm{s}_{1},1)\leq Q(\bm{s}_{1},0)$ and
$V(\bm{s}_{2})=Q(\bm{s}_{2},0)$, then $V(\bm{s}_{2})-V(\bm{s}_{1})=Q(\bm{s}_{2},0)-Q(\bm{s}_{1},1)\geq Q(\bm{s}_{2},0)-Q(\bm{s}_{1},0)\geq\Delta_{2}-\Delta_{1}$.
\end{itemize}
This completes the proof of Lemma \ref{lem:slope wrt aoci}.

\subsection{Proof of Lemma \ref{lem:slope wrt aoi}\label{subsec:Proof-of-slope wrt aoi}}

Let $\bm{s}_{1}=(\Delta,\delta_{1})$ and $\bm{s}_{2}=(\Delta,\delta_{2})$.
We use mathematical induction and the VIA to prove Lemma \ref{lem:slope wrt aoi}.
Specifically, we need to prove that for any $\delta_{1}\leq\delta_{2}$
\begin{equation}
V_{k}(\bm{s}_{1})-V_{k}(\bm{s}_{2})\leq\delta_{2}-\delta_{1},\quad k=0,1,\ldots\label{eq:slope-aoi}
\end{equation}
if $p_{r}(\delta_{1})-p_{r}(\delta_{2})\leq\frac{\delta_{2}-\delta_{1}}{\mathop{V_{k}(\Delta+1,1)}-V_{k}(1,1)}$
for any $k$.

Without loss of generality, we initialize $V_{0}(\bm{s})=0$ for all
$\bm{s}\in\mathcal{S}$. Thus, (\ref{eq:slope-aoi}) holds for $k=0$.
Next, we assume that (\ref{eq:slope-aoi}) holds up till $k>0$ and
we examine whether it holds for $k+1$. 

When $a=0$, we have 
\begin{align}
 & Q_{k}(\bm{s}_{2},0)+(\delta_{2}-\delta_{1})\nonumber \\
= & \Delta+V_{k}(\Delta+1,\delta_{2}+1)+\delta_{2}-\delta_{1}\nonumber \\
\geq & \Delta+V_{k}(\Delta+1,\delta_{1}+1)=Q_{k}(\bm{s}_{1},0).
\end{align}

When $a=1$, we have
\begin{align}
 & Q_{k}(\bm{s}_{2},1)+(\delta_{2}-\delta_{1})\nonumber \\
= & \Delta+\omega C_{u}+p_{s}(1-p_{r}(\delta_{2}))V_{k}(1,1)+p_{f}\mathop{V_{k}(\Delta+1,\delta_{2}+1)}\nonumber \\
 & +p_{s}p_{r}(\delta_{2})\mathop{V_{k}(\Delta+1,1)}+\delta_{2}-\delta_{1}\nonumber \\
= & \Delta+\omega C_{u}+p_{s}V_{k}(1,1)+p_{s}p_{r}(\delta_{2})(V_{k}(\Delta+1,1)-V_{k}(1,1))\nonumber \\
 & +p_{f}\mathop{V_{k}(\Delta+1,\delta_{2}+1)}+\delta_{2}-\delta_{1}\nonumber \\
\overset{(a)}{\geq} & \Delta+\omega C_{u}+p_{s}V_{k}(1,1)+p_{s}(V_{k}(\Delta+1,1)-V_{k}(1,1))\nonumber \\
 & \times\left(p_{r}(\delta_{1})-\frac{\delta_{2}-\delta_{1}}{\mathop{V_{k}(\Delta+1,1)}-V_{k}(1,1)}\right)\nonumber \\
 & +p_{f}(\mathop{V_{k}(\Delta+1,\delta_{1}+1)}-(\delta_{2}-\delta_{1}))+\delta_{2}-\delta_{1}\nonumber \\
= & \Delta+\omega C_{u}+p_{s}V_{k}(1,1)+p_{s}p_{r}(\delta_{1})(V_{k}(\Delta+1,1)-V_{k}(1,1))\nonumber \\
 & +p_{f}(\mathop{V_{k}(\Delta+1,\delta_{1}+1)}+(1-p_{s}-p_{f})(\delta_{2}-\delta_{1})\nonumber \\
= & \Delta+\omega C_{u}+p_{s}V_{k}(1,1)+p_{s}p_{r}(\delta_{1})(V_{k}(\Delta+1,1)-V_{k}(1,1))\nonumber \\
 & +p_{f}(\mathop{V_{k}(\Delta+1,\delta_{1}+1)}\nonumber \\
= & Q_{k}(\bm{s}_{1},1),
\end{align}
where (a) holds if $p_{r}(\delta_{1})-p_{r}(\delta_{2})\leq\frac{\delta_{2}-\delta_{1}}{\mathop{V_{k}(\Delta+1,1)}-V_{k}(1,1)}$.

Since $V_{k+1}({\bf s})=\underset{a}{\min}Q_{k}({\bf s},a)$, we can
show that $V_{k+1}(\bm{s}_{1})-V_{k+1}(\bm{s}_{2})\leq\delta_{2}-\delta_{1}$
if $p_{r}(\delta_{1})-p_{r}(\delta_{2})\leq\frac{\delta_{2}-\delta_{1}}{\mathop{V_{k}(\Delta+1,1)}-V_{k}(1,1)}$
for any $k\in\mathbb{Z}_{\geq0}$ in four cases as follows:
\begin{itemize}
\item Case 1: If $V_{k+1}(\bm{s}_{1})=Q_{k}(\bm{s}_{1},0)$ and $V_{k+1}(\bm{s}_{2})=Q_{k}(\bm{s}_{2},0)$,
then $V_{k+1}(\bm{s}_{1})-V_{k+1}(\bm{s}_{2})=Q_{k}(\bm{s}_{1},0)-Q_{k}(\bm{s}_{2},0)\leq\delta_{2}-\delta_{1}$.
\item Case 2: If $V_{k+1}(\bm{s}_{1})=Q_{k}(\bm{s}_{1},1)$ and $V_{k+1}(\bm{s}_{2})=Q_{k}(\bm{s}_{2},1)$,
then $V_{k+1}(\bm{s}_{1})-V_{k+1}(\bm{s}_{2})=Q_{k}(\bm{s}_{1},1)-Q_{k}(\bm{s}_{2},1)\leq\delta_{2}-\delta_{1}$.
\item Case 3: If $V_{k+1}(\bm{s}_{1})=Q_{k}(\bm{s}_{1},0)\leq Q_{k}(\bm{s}_{1},1)$
and $V_{k+1}(\bm{s}_{2})=Q_{k}(\bm{s}_{2},1)$, then $V_{k+1}(\bm{s}_{1})-V_{k+1}(\bm{s}_{2})=Q_{k}(\bm{s}_{1},0)-Q_{k}(\bm{s}_{2},1)\leq Q_{k}(\bm{s}_{1},1)-Q_{k}(\bm{s}_{2},1)\leq\delta_{2}-\delta_{1}$.
\item Case 4: If $V_{k+1}(\bm{s}_{1})=Q_{k}(\bm{s}_{1},1)\leq Q_{k}(\bm{s}_{1},0)$
and $V_{k+1}(\bm{s}_{2})=Q_{k}(\bm{s}_{2},0)$, then $V_{k+1}(\bm{s}_{1})-V_{k+1}(\bm{s}_{2})=Q_{k}(\bm{s}_{1},1)-Q_{k}(\bm{s}_{2},0)\leq Q_{k}(\bm{s}_{1},0)-Q_{k}(\bm{s}_{2},0)\leq\delta_{2}-\delta_{1}$.
\end{itemize}
By taking limits on both sides of (\ref{eq:slope-aoi}) and by (\ref{eq:lim-value-function}),
we complete the proof of Lemma \ref{lem:slope wrt aoi}. 

\subsection{Proof of Theorem \ref{thm:threshold-structure-general} \label{subsec:Proof-of-threshold-structure}}

The optimal updating policy can be obtained by leveraging the VIA.
In particular, we investigate the difference of the state-action value
function. Let $\bm{s}=(\Delta,\delta)$. According to Lemmas \ref{lem:value function monotony wrt AOCI}-\ref{lem:slope wrt aoi},
if $p_{r}(1)-p_{r}(\delta+1)\leq\frac{\delta}{\mathop{V_{k}(\Delta+1,1)}-V_{k}(1,1)}$
for any $k\in\mathbb{Z}_{\geq0}$, we have 
\begin{align}
 & Q_{k}(\bm{s},0)-Q_{k}(\bm{s},1)\nonumber \\
= & p_{s}\left(V_{k}(\Delta+1,\delta+1)-p_{r}(\delta)V_{k}(\Delta+1,1)\right)\nonumber \\
 & -p_{s}(1-p_{r}(\delta))V_{k}(1,1)-\omega C_{u}\nonumber \\
= & p_{s}\left(V_{k}(\Delta+1,\delta+1)-V_{k}(\Delta+1,1)\right)\nonumber \\
 & +p_{s}(1-p_{r}(\delta))\left(V_{k}(\Delta+1,1)-V_{k}(1,1)\right)-\omega C_{u}\nonumber \\
\geq & p_{s}(1-p_{r}(\delta))\Delta-p_{s}\delta-\omega C_{u}.
\end{align}
We can see that $Q_{k}(\bm{s},0)-Q_{k}(\bm{s},1)$ is lower bounded
by $p_{s}(1-p_{r}(\delta))\Delta-p_{s}\delta-\omega C_{u}$, which
is the sum of an increasing positive function with respect to $\Delta$
and two negative constants.  It is evident that there exists a positive
integer $\underline{\Delta}(\delta)$ such that $\underline{\Delta}(\delta)$
is the minimum value satisfying $p_{s}(1-p_{r}(\delta))\Delta-p_{s}\delta-\omega C_{u}\geq0$.
Therefore, if $\Delta\geq\underline{\Delta}(\delta)$, then we have
$Q_{k}(\bm{s},0)-Q_{k}(\bm{s},1)\geq0$. According to the definition
of the optimal policy in (\ref{eq:optimal-policy-Q}), the optimal
policy for a given $\delta$ is to update when $\Delta\geq\underline{\Delta}(\delta)$. 

\subsection{Proof of Lemma \ref{lem:lemma2}\label{subsec:Proof-of-Lemma 2}}

The proof is similar to that of Lemma \ref{lem:value function monotony wrt AOCI}.
By initializing $V_{0}(s)=0$ for all $s\in\mathcal{S}$, it is easy
to see that $V_{k}(s_{1})\leq V_{k}(s_{2})$ holds for $k=0$. Next,
we assume that $V_{k}(s_{1})\leq V_{k}(s_{2})$ holds up till $k>0$
and we examine whether it holds for $k+1$. 

When $a=0$, we have $Q_{k}(s_{1},0)=s_{1}+\mathop{V_{k}(s_{1}+1)}$
and $Q_{k}(s_{2},0)=s_{2}+\mathop{V_{k}(s_{2}+1)}$. Since $s_{1}\leq s_{2}$
and $V_{k}(s_{1})\leq V_{k}(s_{2})$, we can easily see that $Q_{k}(s_{1},0)\leq Q_{k}(s_{2},0)$.

When $a=1$, the state-action value functions at iteration $k$ are
given by 
\begin{align}
Q_{k}(s_{1},1)= & s_{1}+\omega C_{u}+p_{s}(1-p_{r})V_{k}(1)\nonumber \\
 & +(p_{f}+p_{s}p_{r})\mathop{V_{k}(s_{1}+1)},
\end{align}
 and 
\begin{align}
Q_{k}(s_{2},1)= & s_{2}+\omega C_{u}+p_{s}(1-p_{r})V_{k}(1)\nonumber \\
 & +(p_{f}+p_{s}p_{r})\mathop{V_{k}(s_{2}+1)}.
\end{align}
 Bearing in mind that $V_{k}(s_{1})\leq V_{k}(s_{2})$, we can also
verify that $Q_{k}(s_{1},1)\leq Q_{k}(s_{2},1)$.

Altogether, we can assert that $V_{k+1}(s_{1})\leq V_{k+1}(s_{2})$
for any $k$. By taking limits on both sides of $V_{k}(s_{1})\leq V_{k}(s_{2})$
and by (\ref{eq:lim-value-function}), we complete the proof of Lemma
\ref{lem:lemma2}. 

\subsection{Proof of Theorem \ref{thm:threshold type} \label{subsec:Proof-of-Theorem-1}}

Suppose $\pi^{*}(s_{1})=1$, we have $Q(s_{1},0)-Q(s_{1},1)\geq0$.
Therefore, the optimal updating policy has a threshold structure if
$Q(s,a)$ has a sub-modular structure, that is, 
\begin{equation}
Q(s_{1},0)-Q(s_{1},1)\leq Q(s_{2},0)-Q(s_{2},1),\label{eq:sub-modular}
\end{equation}
for any $s_{1},s_{2}\in\mathcal{S}$ and $s_{1}\leq s_{2}$.

According to the definition of $Q(s,a)$, we have 
\begin{align}
Q(s_{1},0)-Q(s_{1},1)= & p_{s}(1-p_{r})[V(s_{1}+1)-V(1)]-\omega C_{u}
\end{align}
and 
\begin{align}
Q(s_{2},0)-Q(s_{2},1)= & p_{s}(1-p_{r})[V(s_{2}+1)-V(1)]-\omega C_{u}.
\end{align}

Since $V\left(s_{1}+1\right)\leq V(s_{2}+1)$, it is easy to see that
(\ref{eq:sub-modular}) holds. Along with $Q(s_{1},0)-Q(s_{1},1)\geq0$,
we complete the proof of Theorem \ref{thm:threshold type}.

\subsection{Proof of Lemma \ref{lem:total average cost}\label{subsec:Proof-of-expected value of average cost}}

\begin{figure}[t]
\centering

\includegraphics[width=0.5\textwidth]{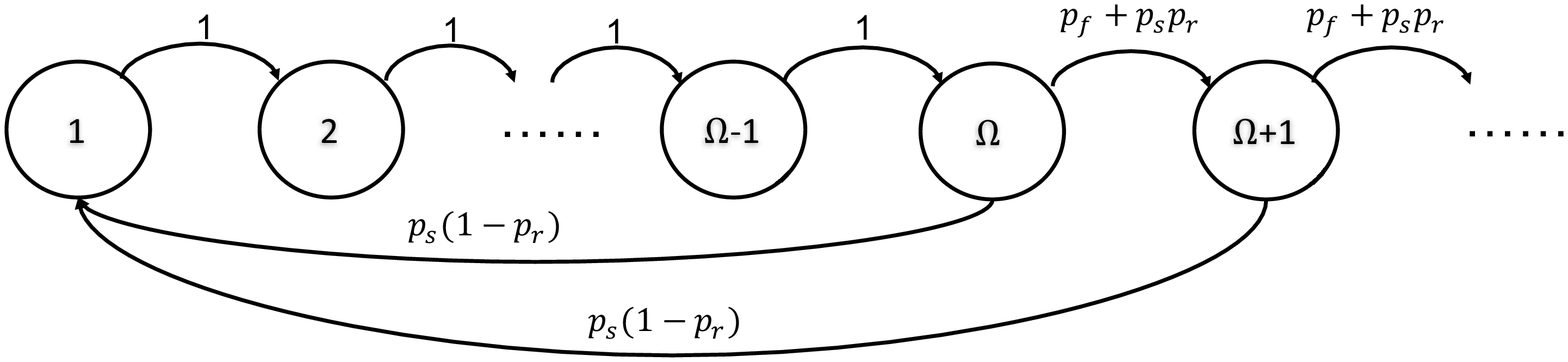}\caption{\label{fig:probability of stability}The states transitions under
a threshold policy with the threshold of $\Omega$.}
\end{figure}

When $\hat{\Delta}=\infty$, for any threshold policy with the threshold
of $\Omega$, the MDP can be modeled through a Discrete Time Markov
Chain (DTMC) with the same states, which is illustrated in Fig. \ref{fig:probability of stability}.
Let $\varphi_{s}$ denote the steady state probability of state $s$.
The balance equations of the DTMC are given as follows:
\begin{equation}
\begin{cases}
\varphi_{s}=\varphi_{s-1}, & 2\leq s\leq\Omega,\\
\varphi_{s}=p_{z}\varphi_{s-1}, & s>\Omega,
\end{cases}
\end{equation}
where $p_{z}=p_{f}+p_{s}p_{r}$. Then, the steady-state probability
of the DTMC can be expressed with $\varphi_{1}$. Specifically, 
\begin{equation}
\varphi_{s}=\begin{cases}
\varphi_{1}, & \text{if }2\leq s\leq\Omega,\\
\varphi_{1}p_{z}^{s-\Omega}, & \text{otherwise}.
\end{cases}\label{eq: phis}
\end{equation}
Since $\sum_{s=1}^{\infty}\varphi_{s}=1$, we can derive $\varphi_{1}=\frac{1-p_{z}}{\Omega(1-p_{z})+p_{z}}$.
By substitute $\varphi_{1}$ into Eq. (\ref{eq: phis}), we can obtain
the closed-form of the steady-state probability, which is given by
\[
\varphi_{s}=\begin{cases}
\frac{1-p_{z}}{\Omega(1-p_{z})+p_{z}}, & \text{if }2\leq s\leq\Omega,\\
\frac{(1-p_{z})p_{z}^{s-\Omega}}{\Omega(1-p_{z})+p_{z}}, & \text{otherwise}.
\end{cases}
\]

Then, the average cost under the threshold policy can be computed
as $J(\Omega)=J_{1}(\Omega)+J_{2}(\Omega)$, where 
\begin{align}
J_{1}(\Omega) & =\stackrel[s=1]{\infty}{\sum}\varphi_{s}s\nonumber \\
 & =\frac{1-p_{z}}{\Omega(1-p_{z})+p_{z}}\left(\frac{\Omega(\Omega-1)}{2}+\frac{\Omega}{1-p_{z}}+\frac{p_{z}}{(1-p_{z})^{2}}\right),
\end{align}
and
\begin{equation}
J_{2}(\Omega)=\sum_{s=\Omega}^{\infty}\varphi_{s}\omega C_{u}=\frac{\omega C_{u}}{\Omega(1-p_{z})+p_{z}}.
\end{equation}

This completes the proof of Lemma \ref{lem:total average cost}.

\subsection{Proof of Theorem \ref{thm:closed-form} \label{subsec:Proof-of-closed-form}}

According to Lemma \ref{lem:total average cost}, we have 
\begin{align}
J(\Omega)= & J_{1}(\Omega)+J_{2}(\Omega)\nonumber \\
= & \frac{1-p_{z}}{\Omega(1-p_{z})+p_{z}}\left(\frac{\Omega^{2}-\Omega}{2}+\frac{\Omega+\omega C_{u}}{1-p_{z}}+\frac{p_{z}}{(1-p_{z})^{2}}\right).
\end{align}
We derive the optimal threshold $\Omega^{*}$ by relaxing $\Omega$
to a continuous variable. We first calculate the second order derivative
of $J(\Omega)$ as follow,
\begin{align}
\frac{\partial^{2}J(\Omega)}{\partial\Omega}= & \frac{1-p_{z}}{\Omega(1-p_{z})+p_{z}}-2\left[\frac{1-p_{z}}{\Omega(1-p_{z})+p_{z}}\right]^{2}\nonumber \\
 & \times\left(\frac{\Omega^{2}-\Omega}{2}+\frac{\Omega+\omega C_{u}}{1-p_{z}}+\frac{p_{z}}{(1-p_{z})^{2}}\right)\nonumber \\
 & +2\left[\frac{1-p_{z}}{\Omega(1-p_{z})+p_{z}}\right]^{3}\nonumber \\
 & \times\left(\frac{\Omega^{2}-\Omega}{2}+\frac{\Omega+\omega C_{u}}{1-p_{z}}+\frac{p_{z}}{(1-p_{z})^{2}}\right)\nonumber \\
= & \frac{(1-p_{z})\left[p_{z}+2\omega C_{u}(1-p_{z})\right]}{\left[\Omega(1-p_{z})+p_{z}\right]^{3}},
\end{align}
where $p_{z}=p_{f}+p_{s}p_{r}\leq1$. Since $\frac{\partial^{2}J(\Omega)}{\partial\Omega}\geq0$,
$J(\Omega)$ is a convex function with respect to $\Omega$. Then,
we calculate the first order derivative of $J(\Omega)$ as follow,
\begin{align}
\frac{\partial J(\Omega)}{\partial\Omega} & =\frac{1-p_{z}}{\Omega(1-p_{z})+p_{z}}\left(\frac{2\Omega-1}{2}+\frac{1}{1-p_{z}}\right)\nonumber \\
 & -\left[\frac{1-p_{z}}{\Omega(1-p_{z})+p_{z}}\right]^{2}\nonumber \\
 & \times\left(\frac{\Omega^{2}-\Omega}{2}+\frac{\Omega+\omega C_{u}}{1-p_{z}}+\frac{p_{z}}{(1-p_{z})^{2}}\right).
\end{align}
The optimal threshold can be obtained by setting $\frac{\partial J(\Omega)}{\partial\Omega}$
to zero. The solution to $\frac{\partial J(\Omega)}{\partial\Omega}=0$
is 
\begin{equation}
\Omega'=\frac{\sqrt{p_{z}+2\omega C_{u}(1-p_{z})}-p_{z}}{1-p_{z}}.
\end{equation}
Since $\Omega^{'}$ may not be an integer, the optimal threshold can
be expressed as 
\begin{equation}
\Omega^{*}=\arg\min(J(\left\lfloor \Omega'\right\rfloor ),J(\left\lceil \Omega'\right\rceil )).\label{eq: threshold-optimal}
\end{equation}

\subsection{Proof of Corollary \ref{cor:optimal-threshold}\label{subsec:Proof-of-Corollary1}}

The  properties of $\Omega^{*}$ in terms of $C_{u}$, $p_{s}$, and
$M$ can be proved by analyzing $\Omega'$, which is given by 
\begin{equation}
\Omega'=\frac{\sqrt{p_{z}+2\omega C_{u}(1-p_{z})}-p_{z}}{1-p_{z}}.\label{eq:threshold}
\end{equation}

\begin{itemize}
\item According to Eq. (\ref{eq:threshold}), it is easy to see that $\Omega'$
is an increasing function of $C_{u}$. Then, $\Omega^{*}$ is a non-decreasing
function of $C_{u}$ due to rounding. 
\item To analyze the relationship between $\Omega'$ and $p_{s}$, we first
investigate how $\Omega'$ varies with $p_{z}$. Particularly, we
calculate the derivative of $\Omega'$ with respect to $p_{z}$ as
follow,
\begin{align}
 & \frac{\partial\Omega^{\prime}}{\partial p_{z}}\nonumber \\
= & \frac{(1-p_{z})\left[\frac{1}{2}\left[p_{z}+2\omega C_{u}(1-p_{z})\right]^{-\frac{1}{2}}(1-2\omega C_{u})-1\right]}{(1-p_{z})^{2}}\nonumber \\
 & +\frac{\sqrt{p_{z}+2\omega C_{u}(1-p_{z})}-p_{z}}{(1-p_{z})^{2}}\nonumber \\
= & \frac{(1-p_{z})\left[\frac{1}{2}\left[p_{z}+2\omega C_{u}(1-p_{z})\right]^{-\frac{1}{2}}(1-2\omega C_{u})\right]}{(1-p_{z})^{2}}\nonumber \\
 & +\frac{-1+\sqrt{p_{z}+2\omega C_{u}(1-p_{z})}}{(1-p_{z})^{2}}\nonumber \\
= & \frac{\frac{1}{2}(1-p_{z})(1-2\omega C_{u})-\left[p_{z}+2\omega C_{u}(1-p_{z})\right]^{\frac{1}{2}}}{\left[p_{z}+2\omega C_{u}(1-p_{z})\right]^{\frac{1}{2}}(1-p_{z})^{2}}\nonumber \\
 & +\frac{p_{z}+2\omega C_{u}(1-p_{z})}{\left[p_{z}+2\omega C_{u}(1-p_{z})\right]^{\frac{1}{2}}(1-p_{z})^{2}}\nonumber \\
= & \frac{\frac{1}{2}(1-p_{z})(1+2\omega C_{u})+p_{z}-\left[p_{z}+2\omega C_{u}(1-p_{z})\right]^{\frac{1}{2}}}{\left[p_{z}+2\omega C_{u}(1-p_{z})\right]^{\frac{1}{2}}(1-p_{z})^{2}}\nonumber \\
= & \frac{A-B}{\left[p_{z}+2\omega C_{u}(1-p_{z})\right]^{\frac{1}{2}}(1-p_{z})^{2}},
\end{align}
where $A=\frac{1}{2}(1-p_{z})(1+2\omega C_{u})+p_{z}$ and $B=\left[p_{z}+2\omega C_{u}(1-p_{z})\right]^{\frac{1}{2}}$.
Since both $A$ and $B$ are positive, we show that $A\geq B$ by
comparing $A^{2}$ and $B^{2}$. Specifically,
\begin{align}
A^{2}-B^{2}= & \frac{1}{4}(1-p_{z})^{2}(1+2\omega C_{u})^{2}+p_{z}^{2}\nonumber \\
 & +(1-p_{z})(1+2C_{u})p_{z}-p_{z}-2\omega C_{u}(1-p_{z})\nonumber \\
= & \frac{1}{4}(1-p_{z})^{2}(1+2\omega C_{u})^{2}-2\omega C_{u}(1-p_{z})^{2}\nonumber \\
= & (1-p_{z})^{2}\left(\omega C_{u}-\frac{1}{2}\right)^{2}\geq0.
\end{align}
Hence, $\Omega'$ is a non-decreasing function of $p_{z}$. Furthermore,
$p_{z}=p_{f}+p_{s}p_{r}=1-p_{s}+\frac{p_{s}}{M}=1-\frac{M-1}{M}p_{s}$
is a decreasing function of $p_{s}$ and $M$. Therefore, $\Omega'$
is non-increasing of $p_{s}$ and $M$. According to Eq. (\ref{eq: threshold-optimal}),
$\Omega^{*}$ is a non-increasing function of $p_{s}$ and $M$.
\end{itemize}
\bibliographystyle{IEEEtran}
\bibliography{AoI}

\end{document}